\documentclass[12pt]{article}
\usepackage[a4paper, total={6in, 8in}]{geometry}
\usepackage{graphicx}
\usepackage{amsmath, amssymb, amsfonts}
\usepackage{mathrsfs}
\usepackage{multirow}
\usepackage[figuresright]{rotating}
\usepackage[title]{appendix}
\usepackage{xcolor}
\usepackage{textcomp}
\usepackage{booktabs}
\usepackage{algorithm}
\usepackage{algorithmicx}
\usepackage{algpseudocode}
\usepackage{graphicx}
\usepackage{xcolor,colortbl}
\usepackage{soul}
\usepackage{graphicx}
\usepackage[utf8]{inputenc}
\usepackage{ulem}
\usepackage[T1]{fontenc}
\usepackage{lscape}
\usepackage{pifont}
\usepackage{footnote}
\newcommand{\cmark}{\ding{51}}%
\newcommand{\xmark}{\ding{55} }%

\usepackage{hyperref}
\usepackage{listings}
\usepackage{authblk}
\usepackage{lineno}
% ...
\PassOptionsToPackage{hyphens}{url}\usepackage{hyperref}
\usepackage{xurl}
\usepackage{setspace}
\begin{document}

\section*{Self-Archiving Statement}
This is the accepted version of the following article, which has been published in final form in \textit{Nature Machine Intelligence}:\\

\hspace{-20pt} A. Tashakori, et al., "Capturing complex hand movements and object interactions using machine learning-powered stretchable smart textile gloves" \textit{Nature Machine Intelligence}, vol. 6, no. 1, pp. 106–118, 2024. 

\hspace{-20pt} DOI: \href{https://doi.org/10.1038/s42256-023-00780-9}{https://doi.org/10.1038/s42256-023-00780-9}.\\

\hspace{-20pt} For the final published version, please refer to the journal's official website:

\hspace{-20pt} \href{https://www.nature.com/articles/s42256-023-00780-9}{https://www.nature.com/articles/s42256-023-00780-9}.\\

\newpage

\title{Capturing Complex Hand Movements and Object Interactions Using Machine Learning Powered Stretchable Smart Textile Gloves}

\author[1, 2, *]{\small Arvin Tashakori}
\author[2, 3]{\small Zenan Jiang}
\author[2]{\small Amir Servati}
\author[2]{\small Saeid Soltanian}
\author[2, 3]{\small Harishkumar Narayana}
\author[2, 3]{\small Katherine Le}
\author[2]{\small Caroline Nakayama}
\author[4, 5]{\small Chieh-ling Yang}
\author[1]{\small Z Jane Wang}
\author[6, 7]{\small Janice J Eng}
\author[1, 2, *]{\small Peyman Servati}

\affil[1]{\footnotesize Department of Electrical and Computer Engineering, University of British Columbia, Vancouver BC, Canada, V6T 1Z4}
\affil[2]{\footnotesize Texavie Technologies Inc., 148-970 Burrard St, Vancouver BC, Canada, V6Z 2R4}
\affil[3]{\footnotesize Department of Materials Engineering, University of British Columbia, Vancouver BC, Canada, V6T 1Z4}
\affil[4]{\footnotesize Department of Occupational Therapy and Graduate Institute of Behavioral Sciences, College of Medicine, Chang Gung University, Taoyuan City, Taiwan}
\affil[5]{\footnotesize Department of Physical Medicine and Rehabilitation, Chang Gung Memorial Hospital, Chiayi, Taiwan}
\affil[6]{\footnotesize Department of Physical Therapy, Faculty of Medicine, University of British Columbia, Vancouver BC, Canada, V6T 1Z3}
\affil[7]{\footnotesize Centre for Hip Health and Mobility, Vancouver Coastal Health Research Institute, Vancouver BC, Canada, V5Z 1M9}
\affil[*]{Corresponding authors: \href{mailto:arvin@ece.ubc.ca}{arvin@ece.ubc.ca}; \href{mailto:peymans@ece.ubc.ca}{peymans@ece.ubc.ca}}

\date{} 

\maketitle

\begin{abstract}
Accurate real-time tracking of dexterous hand movements and interactions has numerous applications in human-computer interaction, metaverse, robotics, and tele-health. Capturing realistic hand movements is challenging because of the large number of articulations and degrees of freedom. Here, we report accurate and dynamic tracking of articulated hand and finger movements using stretchable, washable smart gloves with embedded helical sensor yarns and inertial measurement units. The sensor yarns have a high dynamic range, \textcolor{black}{responding to low 0.005 \% to high 155 \% strains}, and show stability during extensive use and washing cycles. We use multi-stage machine learning to report average joint angle estimation root mean square errors of 1.21 and 1.45 degrees for intra- and inter-subjects cross-validation, respectively, matching accuracy of costly motion capture cameras without occlusion or field of view limitations. We report a data augmentation technique that enhances robustness to noise and variations of sensors. We demonstrate accurate tracking of dexterous hand movements during object interactions, opening new avenues of applications including accurate typing on a mock paper keyboard, recognition of complex dynamic and static gestures adapted from American Sign Language and object identification.
\end{abstract}

\section{Introduction}\label{sec1}

Real-time tracking of hand movements \cite{Luo2021,moin2021,Zhou2020,Chen2020,Glauser2019,wang2020gesture,zhu2022soft,hughes2020simple,sun2022augmented,sundaram2019learning} has significant applications in human-computer interaction (HCI), electronic gaming, metaverse and augmented reality (AR) \cite{Chen2020,Lei2019,Wen2021}, rehabilitation \cite{Chen2020,Henderson2021,Hughes2018,Atiqur2019,yang2023perspectives}, sports training, robotics, and tele-surgical applications \cite{Chen2020,Lei2019}. Fueled by recent advances in machine learning (ML) and flexible electronics \cite{Luo2021,moin2021,Zhou2020,Chen2020,Glauser2019,Wen2021,Moin2018}, significant progress has been reported for tracking or gesture recognition using both computer vision (CV) \cite{Chen2020,Karunratanakul2020,Smith2020,xie2021physics,Jiang2019,Guzov2021,Ehsani2020,ge20193d,Wu2020,Kocabas2020,Hu2020,Cao2021} and wearable technologies \cite{moin2021,Zhou2020,Chen2020,Glauser2019,Wen2021,Henderson2021,Hughes2018,Moin2018,Si2022,Zhang2021,Wang2022,dengsen,Pan2021,Liu2021,Denz2021,Science2020,cote2019deep,zhang2023intelligent}. Fixed and costly motion capture camera systems with markers are often used as the gold standard for detailed and articulated hand and finger tracking. CV-based solutions using one or more cameras either attached to the user's headset \cite{wang2020gesture,Atiqur2019,Hu2020} or placed in a specific location \cite{Atiqur2019,Cao2021} are demonstrated as lower-cost consumer solutions. Both motion capture and CV-based technologies are spatially limited to the field of view of cameras and face major challenges due to occlusion by objects, hands or other body parts, poor lighting, and background noise \cite{Chen2020,Lei2019,Atiqur2019}. Current wearable technologies are primarily used for gesture recognition in form of gloves \cite{Luo2021,Zhou2020,Glauser2019,Wen2021,luo2021intelligent}, wrist bands \cite{Wu2020}, or arm sleeves \cite{moin2021,Moin2018}. Researchers have reported integration of different sensors, including surface electromyography (sEMG) electrodes \cite{moin2021,Moin2018}, pressure sensors \cite{Zhou2020,Glauser2019,wang2020gesture,zhu2022soft, hughes2020simple, sun2022augmented,sundaram2019learning,zhou2022scalable}, inertial measurement units (IMUs) \cite{Chen2020}, magnetic sensors \cite{Chen2020}, and optical sensors \cite{Cao2021,Brahmbhatt2020}. However, most of these wearable devices are used only to detect specific gestures with limited accuracy and have not addressed challenges related to reliability, accuracy, and washability of the device \cite{Luo2021,Chen2020}. Current solutions have strict requirements for placement of sensors directly on user's hand and do not address variations in electrical and mechanical properties of the sensors and fit to the users \cite{Luo2021,moin2021,Chen2020,Moin2018}. These factors in addition to lack of washing or sanitization methods lead to limited practical usability and accuracy of these solutions \cite{Luo2021,Chen2020,Lei2020}. 

In this work, we report for the first time a dynamic and accurate dynamic tracking of hand movements, articulated for all finger and wrist joints, using stretchable, wireless and washable smart textile gloves embedded with stretchable helical sensor yarns (HSYs), IMUs and stretchy interconnects. Using our multi-stage ML algorithms, we report average joint angle estimation root mean square error (RMSE) of 1.21 and 1.45 degrees for intra- and inter-subjects cross-validation, respectively, going well beyond the accuracy of published wearable devices and CV systems \cite{Luo2021,moin2021,Zhou2020,Chen2020,Glauser2019}. Our results indicate that the proposed smart gloves and ML algorithms provide a tool for learning dexterous hand and finger movements that goes beyond conventional gesture recognition and compete in accuracy with costly motion capture systems without limitations of field of view, long setup times (e.g., camera calibration time, marker setup time) and sensitivity to occlusion and image noise that are highly prevalent in hand tracking applications. Indebted to high dynamic range and reliability of the HSYs in response to stretches and pressures at the tip of the fingers as well as our ML algorithms, we demonstrate reliable tracking of complex hand and finger movements during interaction with objects, which is not practical in camera systems due to occlusion by objects and fingers. We also demonstrate a data augmentation technique that increases robustness of our results by two times in the presence of sizable variations in the performance of the sensors and their fit to the subject. Based on these results, we demonstrate complex potential applications such as dynamic tracking of hand and finger movements, accurate (97.80 \%) typing on a mock paper keyboard, highly accurate dynamic gesture recognition (94.05 \% and 97.31 \% inter- and intra-subject cross-validation accuracy, respectively, for 50 gestures), static gesture recognition (94.60 \% and 97.81 \% inter- and intra-subject cross-validation accuracy, respectively, for 48 gestures) as well as object recognition from grasp patterns (90.20 \% and 95.02 \% inter- and intra-subject cross-validation accuracy, respectively, for 34 objects). Table 1 summarizes overall performance parameters of our system compared to other published works. We believe our stretchable smart textile glove and ML algorithms can unlock new avenues in HCI, movement and therapy assessment in remote health as well as applications in animation and metaverse for learning dexterous human hand functions and interactions.

\section{Smart Textile Gloves}\label{sec2}
Fig. 1a demonstrates the main functionality of smart textile glove, including joint angle estimation, and detecting the grasp pressure in interaction with objects. Fig. 1b illustrates the photograph of our glove with a schematic overlay showing the embedded helical sensor yarns (HSYs), specifically located at each finger joints, tips of each fingers, palm and thumb joints. Wavy 3D stretchable interconnects connecting all sensors to a wireless processing board with a rechargeable battery as well as two 9-axis degrees of freedom (DOF) IMUs are integrated on the dorsal side of the hand and within the textile on top of the forearm. As discussed in the next section, the HSYs are integrated on the surface of a stretchable internal glove fabric and connected using wavy insulated stretchable interconnects. The gloves are then covered with another layer of stretchable fabric and sewn together to provide reliable, durable and accurate operation. The HSYs can detect local deformations including stretches and pressures in the fabric caused by the movements of joints and fingers or forces when interacting with objects or when pressed against one's own hand tissue (Fig. 1c). The two 9-DOF IMUs enable highly accurate tracking of wrist joint quaternion angles.

Fig. 1d shows the schematic diagram of the system model. We used a custom-made wireless processing board (Texavie) that consists of amplifiers, multiplexers, analog to digital converters (ADCs), microprocessors and Bluetooth low-energy (BLE) wireless transmission modules that interact with all the sensors and IMUs and transmit data to an external receiver. We use an iOS mobile app (Texavie) on iPhone or iPad (Apple), or a PC as a data gate and use an ML data pipeline that performs all data processing and ML models, including GlovePose model, 3D visualization (Fig. 1e), contact detection, gesture and object detection algorithms. An ML-based hand joint angle estimation algorithm is developed as GlovePoseML model, which was validated against a motion capture system, as explained in the next sections. 

\section{Helical Sensor Yarns}\label{sec3}

Fig. 2a illustrates the structure of the HSY that are critical to accurate performance of the smart glove for detection of finger and hand movements used in the machine learning models. The stretchable HSYs consist of an elastic core yarn wrapped with metal-coated nanofibers (NFs) in helical form as shown in the figure. The HSYs structure is completed with a final protective coating of polydimethylsiloxane (PDMS) and silicone as the outer shell. PDMS can penetrate deeply into the NF mesh and bond the NFs together during bending, pressing, and stretching providing high durability and dynamic range for the piezoresistive HSYs. Silicone elastomer is then applied over the yarn to attach the sensors to the fabric substrate and to protect the PDMS layer as well as providing stable electric insulation during washing. All the processing steps are scalable to roll-to-roll manufacturing including roll-to-roll sputtering for economically viable fabrication. Fig. 2b and 2c demonstrate the structure of a typical HSY before and after PDMS and elastomer coatings. The NFs form a helical porous layer around the elastic core yarn. With the PDMS matrix binding the fibers together, the contact area of the metalized NFs changes during the external stretching/pressing-releasing cycles, resulting in a change of resistance in response to strain \cite{soltanian2015highly,soltanian2013highly}. The helical structure of the NFs ensures that the HSYs maintain linear responses to a wide range of strains and stresses. Low-temperature curing of PDMS layer ($\sim120\, \mu m$) was employed for higher tensile strength and stretchability \cite{konku2020curing}, while the elastomer shell further improves the tensile strength of the yarns.  We conducted multiple experiments on different parameters of PDMS coating, such as thickness, curing time and temperature, different shapes and materials of mold to achieve optimum HSY dynamic range, sensitivity and durability. We found that an optimized thickness of 120-200 $\mu$m PDMS thickness and 40-60 $\mu$m core HSY provides optimum performance. The HSYs can reach \textcolor{black}{up to 155 \% }stretchability and ultra-flexibility as demonstrated in Fig. 2c.

Fig. 2d depicts the sensor response, defined as change of the yarn resistance to its original value, for the HSYs to a uniaxial strain ranging \textcolor{black}{from 0.005 to 155 \%} at a frequency of 1 Hz, highlighting the exceptional dynamic range and uniformity of the HSYs. As the in-set shows, the HSYs can accurately respond to external strains as low \textcolor{black}{as 0.005 \%} , which outperforms the performance of other published wearable sensors to the best of our knowledge, while still maintaining the response at \textcolor{black}{155 \%} strain \cite{liu2022functionalized}. In addition, the HSYs respond to strains along both longitudinal and vertical directions. Compressive pressure tests on the sensors are illustrated in Extended Data Figure 1c and demonstrate that the sensors can respond to pressure levels as low as 1.7 kPa and up to 1.2 MPa. The exceptional dynamic range, \textcolor{black}{sensitivity, low hysteresis, high linearity, and reliability of the HSYs are key for achieving accurate response in smart glove and tracking of small joint movements, where the stretch in glove fabric is expected to be less than 120 \%. Fig. 2e displays the time-dependent response of the HSYs to uniaxial straining cycles with different maximum magnitude of 1 \%, 2 \%, 5 \%, 10 \%, 20 \%, 40 \%, 80 \%, and 140 \% }. The dependence of the output signals on the frequency of straining cycle is shown in Fig. 2f. The HSY sensors demonstrate accurate sensing performance under bending radius of down to 2.5 mm and repeated pressure from sharp objects like 0.5 mm stainless steel plate and a 0.6 mm diameter plastic pipette tip up to about 2.1 MPa.

The response time of the HSYs to external strain stimuli is also characterized to evaluate the response time of the system. Fig. 2g shows exceptional synchronization of the stimuli and output signals illustrating the possibility for real-time dynamic motion tracking. The durability of the sensors was studied by conducting a 14-hour long overnight test at the strain level of 10 \% and 0.7 Hz frequency (Fig. 2h). Sensitivity of the response to water is also demonstrated in Fig. 2i comparing the response in air and when fully immersed in water. The sensors show less than 10 \% degradation in sensitivity over the entire range of strains. The sensors are integrated on the fabric and stability and washability testing was undertaken over various day-by-day laundry tests, as shown in Fig. 2j. The sensors were first immersed in water, at various temperatures for different times ranging from 5 to 50 minutes with magnetic stirring at different speeds (80, 160 or 400 rpm). The fabric and sensors were dried in air completely before the measurements of sensitivity. Similarly, detergent (dtg) and softener (sft) were added to the water, and the response of the sensors was recorded after they were dried. The sensors show excellent stability after the simulated drying cycles. In the end, we conducted real washing/drying tests by putting the sensors into the laundry machines together with a full load of other clothes, to study the performance of the sensors in actual daily usage scenarios, demonstrating stability during laundry cycles. HSYs demostrate superior specifications to state-of-the art strain sensors, including broad dynamic range \cite{zhou2022scalable,ho2019multifunctional, vu2020highly}, fast response time \cite{tang2018highly, gao20193d, cai2020highly}, durability in extensive cycling \cite{liu2022functionalized}, and excellent washability \cite{zhou2022scalable, xu2020moisture}. In addition to washability for HSY embedded fabric, the enclosure box used for PCB hardware is designed with embedded custom made waterproofing rubber flanges that blocks water leakage in normal washing conditions and the PCB is coated with a protective coating and placed in a plastic pouch. This makes the entire glove system washable after removal of the rechargeable battery, tested for over tens of cycles of washing and drying.

\section{Machine Learning Model}\label{sec31}

In this work, we developed an ML model that can dynamically estimate joint angles for all finger joints and wrist with high accuracy. This ML model (GlovePoseML) is the core of the algorithm which is supplemented with other neural network-based models in the output to tune the response for specific applications and demonstrations such as keyboard and object detection. To develop and train GlovePoseML model, we collected a large dataset (over 3,000,000 frames with a sampling rate of 20Hz) from our smart glove with attached markers for gold standard motions capture system from five participants with different hand sizes. The participants perform complex hand transient movements collected for various tasks including random finger movements, grabbing objects and switching between different gestures (e.g., making fist or paper). As shown in Fig. 3f, our GlovePoseML model consists of a 2 layer stacked recurrent neural-network-based bi-directional long-short-term-memory (Bi-LSTM) followed by 2 fully-connected (FC) layers and two activation layers. We use this regression architecture for estimating hand joint angles and tactile information from incoming data. Our model uses a two-second history of normalized sensor and motion capture data as input for training. For inter-subject cross-validation, we select one user's data as the testing dataset and others as the training dataset, repeat this step for all users, and report the average results. For intra-subject, we performed ten-fold cross-validation over each user's data and report the average value. With the core model trained, test data can be fed to the model and estimated joint angles are sent through a WebSocket for 3D visualization using Unity software. Fig. 3a demonstrates the linear plot comparison of angles estimated by our ML model for test data and the motion capture gold standard for different finger joint flexion or abduction (if applicable) and wrist joint flexion, abduction and supination. Fig. 3c and 3d report inter- and intra-subject cross-validation results for R2 and RMSE of joint angles, respectively, displaying the model's accuracy for each joint. More detailed results for the accuracy values can be found in Supplementary Table 1. In particular, the results demonstrate high accuracy of 1.45 (min: 0.69, max: 2.71) and 1.21 (min: 0.36, max: 2.26) degrees RMSE for inter- and intra-subject cross-validation results, respectively. We have demonstrated that our system operates with accuracy in scenarios where current CV methods and wearable devices provide limited or low-accuracy results, such as random finger movement in complex gestures (Fig. 3b-i), visual occlusion by objects (Fig. 3b-ii) and dark conditions (Fig. 3b-iii), shown in detail in our demo video. Such high accuracy values arise from several factors including the stability and dynamic range of our HSYs and reliable positioning of the HSYs on the joints in stretchable smart gloves, leading to similar responses and trends among different subjects and different sessions and accurate performance of our ML algorithm. Our results indicate that our smart glove and ML model can potentially replace high-end, bulky and expensive motion capture systems and is not prone to common issues associated with these systems including losing track of markers due to occlusion, light noise or marker swap.

\section{Data Augmentation Technique}\label{sec4}

In practical applications, the smart gloves and the embedded HSYs have small but unavoidable variations in the values of resistances and stretch sensitivities and how they fit to different subjects or the same subject in different sessions. By drawing inspiration from nature that adapts to changes in external environment, we developed a data augmentation technique to enhance robustness to the small variations of HSYs, their placement and fit to the subjects. As shown in Fig. 3e, the pre-training algorithm uses data augmentation to duplicate (i) the original training data and apply some expected variations to the original data, including (ii) random sensor amplitude scaling or DC shift, (iii) random sensor masking, or (iv) additive noise. Then, we use the augmented dataset to train GlovePoseML model to enhance robustness. Fig. 3g and h demonstrate average R2 and RMSE, respectively, for the original model as well as that incorporating the data augmentation technique. These results highlight that the proposed pre-training is highly effective in improving the accuracy ($\sim 2 \times$ reduction in average RMSE) of the system in the presence of unavoidable variations. This step enables our ML model to capture invariant representations of the movement accurately and also reduces the necessity for excessive calibration and retraining in the case of variations in the performance or fit. By developing this pre-training method (Supplementary Algorithm 1), we significantly improve the robustness of our smart glove in comparison to published work that are typically made in the lab and mostly tested by attaching sensors to joints of subjects in a highly controlled and unrepeatable fashion \cite{Zhou2020,Chen2020}. 

\section{Application Examples}\label{sec5}

We rely on our dexterous hand movements to perform everyday tasks, from manipulating objects, using computer or communicating with each other \cite{Chen2020}. To demonstrate our work's possible applications, we demonstrate six examples of our smart gloves and ML system for capturing complex real-time hand and finger poses and movements, typing on a mock keyboard, tracking movements in air, interactions and grasping of objects, dynamic and static gesture recognition. These applications demonstrate how achieving dynamic accuracy can open up applications for our smart gloves and ML system.

For the first example, we report dynamic (ML response time of $\sim 5 ms$ in processing of data on a regular PC with Intel Core i9-9900K CPU $@$ 3.5 GHz, and 32GB RAM DDR4) articulated tracking of finger movements in complex dynamic gestures as shown in the demonstration video. As seen, the system can follow finger movements and all finger and wrist joints with a high level of accuracy. When the model is trained, the accuracy is not affected in dark conditions or when the hand is grasping a ball which creates visual occlusion for the motion capture cameras making them practically ineffective. This impressive dynamic performance and accuracy is a significant improvement to previously reported works in the literature and can find applications in animation, metaverse and tele-operation (A summary of the performance of our glove system versus other similar systems can be found in Table 1).

For the second example, we demonstrate typing on a mock keyboard printed on a piece of paper laid on a table, which serve as a guide for the user during typing (Fig. 4a). We employed GlovePoseML model trained for both hands and then added two FC layers (Fig. 4c) along with a click detection algorithm running on the response of HSYs at the tip of finger as shown in Fig. 4d. Inter-session cross-validation of the system's accuracy is displayed by different colors in Fig. 4b and shows an average 97.80 \% accuracy for the prediction of the typed letters. As shown in the demonstration video, the accuracy can support complex keyboard functions such as holding shift for capital letters not demonstrated in previous works. Detailed accuracy results can be found in Extended Data Figure 3. 

For the third example, we considered a 3D drawing in the air using the glove as shown in Fig. 4e. In this scenario, users can maneuver over an iOS mobile interface application, as shown in the demonstration video, and use their wrist movements to draw with different colors selected based on which finger pinches the thumb. We reported 2.48, 2.34, and 2.54 degree errors for estimating the angle of wrist flexion or extension (Fig. 4f), abduction or adduction (Fig. 4g), and supination and pronation (Fig. 4h), respectively, when compared with the motion capture system.

For the fourth and fifth examples, we focused on dynamic (Fig. 5a), and static gesture recognition (Fig. 5d). Some gestures, such as those shown in Fig. 5d (static gestures 5, 13, 25, 26, 31, and 32), exhibit high similarity, making it very challenging for current CV algorithms to distinguish. We employed GlovePoseML model for each hand and added two FC layers to map finger joint angles to the list of 50 dynamic gestures or 48 static gestures featuring complex finger and wrist poses. We trained the added two layers while keeping GlovePoseML models as is (Fig. 5j). We report inter- and intra-subject cross-validation results of 94.05 \% and 97.31 \% accuracy for detecting dynamic gestures (More details in Extended Data Figure 4), and inter- and intra-subject cross-validation accuracy of 94.60 \% and 97.81 \%, respectively for 48 static gestures (More details in Extended Data Figure 5). To visualize the effectiveness of our classification algorithm, we concatenated the output from GlovePoseML models for both hands. We employed the t-Distributed Stochastic Neighbor Embedding (t-SNE) method to visualize the data derived from these concatenated outputs (Fig. 5b for dynamic and Fig. 5e for static gestures). Fig. 5c and Fig. 5f show the high sensitivity of our method for each gesture.

For the sixth application, we focused on detecting objects based on the participants' grasp patterns. We employed GlovePoseML model and added two layers of FC neural network, mapping finger joint angles to a list of 34 objects (Fig. 5g) with slight difference in shapes, weights, stiffness, and grasp patterns. Here, the users are instructed to mimic a grasp pattern for each object using an instruction video. We trained these added two layers while keeping the core models as is (Fig. 5j). We report inter- and intra-subject cross-validation results of 90.20 \% and 95.02 \% accuracy in detecting the objects from an individual's grasp, respectively (Fig. 5e, more details in Extended Data Figure 6). The algorithm makes use of different hand poses used for objects (e.g., objects 11 and 34) as well as slight differences in pose and grasp pressure patterns (e.g., objects 9, 10, and 11) for objects with very similar hand pose for accurate recognition. The distribution of clusters representing different objects realized by t-SNE for different subjects in Fig. 5h, as well as object-wise sensitivity of our system reported in Fig. 5i demonstrates effectiveness of our classification method.

\section{Conclusion}\label{sec6}
We report accurate, stretchable, washable, and multi-modal smart textile gloves, with embedded stretchable helical yarn sensors, IMUs and interconnects, which can dynamically track movements of finger and hand joints and grasp forces during object interactions. The lightweight and insulated sensor yarns show high dynamic range in response to strains as low as 0.005 \% and as high as 155 \%, and low hysteresis and high stability during extensive use and washing cycles. Using machine learning algorithms, we demonstrate high accuracy for the smart gloves in estimating hand joint angles with less than 1.21 and 1.45 degrees average RMSE compared to the gold standard motion capture system for intra- and inter-subject cross-validation studies, respectively. Data augmentation technique is developed that enhances robustness of the system to variations in the performance of the sensors, noise, and fit to the subjects. The reported smart gloves and machine learning system highlight the potential for high-accuracy tracking of dexterous hand movements and object interactions without limitations of field of view, occlusion and multi-user challenges associated with camera-based systems. We highlight the performance of our system by demonstrating real-time tracking of complex hand movements with minimal delay, as well as accurate detection of 48 static and 50 dynamic gestures adapted from American Sign Language, typing on a random surface as a mock keyboard, and recognition of objects from grasp pose and forces. These results lay foundation for learning realistic, dexterous human hand movements and object interactions for creating new experiences in metaverse, robotics control, tele-surgery, telerehabilitation, and gaming applications, as well as potential for analysis of muscle function and grasp forces for assessment of patients with hand impairments. By extending the volume of data from different movements, interactions and gestures in various scenarios, increased depth and knowledge of hand dexterous functions will empower future realistic digital experiences and human-robot interactions. 

\section{Methods}\label{sec7}

\subsection{Fabrication}\label{subsec1} 
The HSYs were fabricated using needleless electrospinning (NS LAB Nanospider, Elmarco) for deposition of NFs directly onto a core spandex polyurethane yarn (3600 denier; Weight in grams of 9000 meters of length) with a diameter of 0.5 mm (Crystal Tec, Korea) in a roll-to-roll fashion. Before electrospinning, the core yarn was cleaned using oxygen plasma (Tergio-plus) as well as acetone, isopropanol (IPA), and deionized (DI) water. Polyacrylonitrile (PAN) (Scientific Polymer Products) was dissolved in dimethylformamide (DMF) (10 wt \%) (Fisher) and stirred at 60 $^{\circ}$C for 24 h to form a homogeneous solution for electrospinning under a DC bias of 1.5 kV/cm. The core yarn is held between the bias wires of the electrospinning system and rotated around itself (60 rpm) using an electric motor to form helical coating of the NFs with an average diameter of 300 nm (standard deviation of 25 nm). After NF deposition, the yarns are coated with a ~40-50 nm conformal layer of gold by plasma sputtering (Edward) to form a conductive helical sensor yarn (HSY). Ag-coated Nylon threads (Noble Biomaterials) were used as contact electrodes and knotted to the sensor yarn at desired locations over the length of the yarn and bound with silver paste (Pelco). The Ag-coated nylon thread (Denier:100; linear resistance: 0.8 ohm/cm) is a 3 ply yarn, the single plies are twisted at 16 twists/inch in the S direction and the 3 plies are twisted together at 14 twists/inch in the Z direction. The yarns were cured at 70 $^{\circ}$C for 30 min to reach the optimal mechanical and electrical robustness. The sensor yarns are then encapsulated with PDMS (Dow Corning) by pouring and completely curing at 40 $^{\circ}$C for 24 hours using a tube-shaped mold to form a tubular all-around insulation.

The smart glove was prepared using a stretchable single jersey plated weft-knitted fabric (45 \% Nylon, 45 \% Polyester, and 10 \% Spandex), which was patterned and cut for desired glove patterns. The fabric has the following geometrical and mechanical properties (ASTM-D5035 Standard): stitch density (2142 loops/square inch), thickness (0.61 mm), weight (280 gm/square meter-GSM), breaking force (320 N), breaking elongation (110 \%), and elastic recovery (94.8 \%). The sensor locations and their corresponding wiring were designed for the given size of the glove. A wiring bundle was made for each branch of the sensors routed for each finger (3 - 5 sensors in each branch) by twisting thin (34-38 AWG) insulated copper wires and mounting it on the fabric in a wavy form using a sewing machine to achieve the desired stretchability for the interconnects. The yarn sensors were placed and attached to the designated locations on the fabric corresponding to joint sensors and tips of the finger as shown in Fig. 1a using elastomer (Dragon Skin, Smooth-On\texttrademark). The fabric pieces with sensors are cured at room temperature for 30 min, and then the inside and outside fabrics are sewn together to form the final smart gloves. The interconnects are connected to a flexible printed circuit (FPC) that is plugged into the main board and box. Fabrication process is demonstrated in Extended Data Figure 1a.

\subsection{Characterization of the HSYs}\label{subsec3} 
Tensile testing (INSTRON 5969) was performed to investigate the mechanical performance of the HSYs during stretching \textcolor{black}{at strain >2\%} and pressing. The top end of HSY sample was fixed on the system dynamometer to measure and control the force applied to the sensor, and the other end was anchored onto a steady holder. By adjusting the frequency and amplitude, the output under different strains was recorded. For compression studies, a metal indenter $5\times 5\,mm$ is used to tap on the center of sensors. \textcolor{black}{Low strain tensile testing (<2\%) was conducted using a micro actuator (Zaber Technologies). The current and voltage measurements were acquired simultaneously by a Keysight B1500A semiconductor analyzer.}

To test the durability of the HSYs against daily laundry, the gloves were washed and dried using a typical laundry machine and detergents (Samsung front load, Samsung electric compact dryer, Purex\,\textsuperscript{\tiny\textregistered} liquid laundry detergent, Downy ultra fabric softener liquid). The electrical and mechanical properties of the gloves were monitored after each test. 

\subsection{Electronic Hardware and Software}\label{subsec6}
The movement data is acquired at a sampling rate of 20 Hz using a custom PCB (Texavie) containing multiplexers, pre-amplifiers, and analog-to-digital converters (ADC) to measure 25 HSYs with a 12-bit resolution and two IMUs (Bosch BNO055) controlled by a centralized microprocessor (Nordic Semiconductor) with Bluetooth low energy (BLE) connectivity. The power consumption details are included in Extended Data Figure 2.

Data collection gateway is implemented in a mobile application written in Swift 5 (Texavie), which collects data transmitted from each pair of gloves and stores the data in a database. It also sends a start/stop command through a web socket to a python script running on a personal computer for synchronization with gold-standard motion capture cameras (Optitrack). To compensate the shifting of the baseline resitance, we developed a dynamic baseline correction module in the data acquisition software to subtract the sensors’ resistance values from their average value over a
window of size n = 400 (20 seconds).

\subsection{Data Collection and Training}\label{subsec8}
We collected data from five healthy right-handed participants (two female and three male subjects) between the age of 15 to 35 on both hands. In this study, we used one stretchable glove pair for all the subjects, which provided a conformal fit for the subject during the entire data collection session. Informed consent was obtained from all participants (Ethics obtained from UBC Clinical Research Ethics Board - H21-03021 entitled ``iGRASP:Phase 2 (Version1.0)'').  Our data collection consisted of three main sections collected both from the gold standard motion capture system as well as the pair of proposed gloves system: 1) tracking hand and wrist joint angles in random movements or keyboard typing, 2) dynamic and static hand gesture recognition, and 3) grasping different objects. Each section contains a list of movements that the subjects were asked to follow using a prerecorded video supervising them to do a set of gestures.

\subsubsection{Articulated Dynamic Hand Tracking}\label{subsec9}
We asked participants to follow a set of random finger and wrist movements. To mitigate task-related bias, subjects are asked to follow a set of prerecorded videos. We recorded over 3,000,000 frames of data, including from the HSYs and IMUs for actual finger and wrist joint flexion and supination angles and from the motion capture system. The wrist quaternion angle data is derived from the two IMUs and downsampled to have similar data rate to that of HSY data and then fed as input to GlovePoseML. The decision to fuse both modalities is due to their high correlation with joint angles and similar characteristics and computational efficiency, as treating the data from two sensors in different network branches did not show significant performance improvements.

We developed a recurrent neural network-based regression Bi-LSTM neural network architecture, which maps a 2 seconds window of sensor values to one set of hand joint angles. We used the Adam optimizer with a learning rate of 0.0001. We trained our model for 100 epochs. We used Smooth L1 loss function with $\beta = 0.5$. More comprehensive results on the effect of customizing network architecture on joint-wise RMSE and R2 can be found in Supplementary Data Figures 10 and 11.

To enhance robustness, we augmented collected gloves data using three data transformations: 1) Randomly masking one to three channels, 2) Randomly adding Gaussian noise (with the mean of $\mu = 0$ and standard deviation of $\sigma = 0.06$) to one to three channels, and 3) scaling one to three channels by a random scalar between $0.5$ to $1.5$. Later, we trained our ML model in the multi-task learning setting. More details can be found in Supplementary Algorithm 1. More comprehensive results on the effect of noise level, number of masked or scaled sensors on GlovePoseML trained on normal data and augmented data, in terms of average RMSE and R2 can be found in Supplementary Data Figures 12. A demonstration video showcasing dynamic articulated tracking of finger movements can be found in Supplementary Video 1.

\subsubsection{Keyboard Typing Detection}\label{subsec13}
We asked the users to click on a single button of the keyboard. We developed a Click detection algorithm based on the HSY sensors at the tip of the fingers (more details in Supplementary Algorithm 2. We use two FC layers to map the output of GlovePoseML model to ten keys that the fingers maneuver over using the ten-finger typing method. The keyboard was marked using four motion capture markers placed on the corners of the printed keyboard. This is used to calibrate key locations as our ground truth. At the start of data collection, we asked the participants to pose their hands over the keyboard and stay in this position for 10 seconds as the resting position in the ten-finger typing method. Then, we asked them to type a paragraph containing 100 characters on a printed keyboard for training of the two output layers. We collected five different trials and performed five-fold cross-validation. A demonstration video showcasing typing on a mock paper keyboard can be found in Supplementary Video 2.

\subsubsection{3D Drawing}\label{subsec14}
We developed a Mobile interface app using Swift 5. We developed the cursor's location on the screen based on the relative angle of the two IMUs' quaternion values. Users can choose different colors by pinching the thumb using different fingers detected by using a touch detection algorithm similar to Click detection from the HSYs at the tip of the fingers. A demonstration video showcasing 3D drawing in air can be found in Supplementary Video 3.

\subsubsection{Dynamic and Static Hand Gesture Recognition}\label{subsec11}
We recorded a set of videos each containing one gesture and rest position. The gesture was shown for 5 seconds, and users had 5 seconds to rest between each trial. We collected 20 sets of data for each gesture in total. We trained the last two output layers while freezing GlovePoseML models. Inter-session cross-validation results can be found in Supplementary Data Figures 1, 2, and 3, for dynamic gesture recognition, and Supplementary Data Figures 4, 5, and 6 for static gesture recognition. Also, demonstration videos showcasing static and dynamic hand gesture recognition can be found in Supplementary Videos 4 and 5.

\subsubsection{Object Detection From the Grasp}\label{subsec12}
We recorded a set of videos containing grasping one object and resting position. Grasps were shown for 5 seconds, and users had 5 seconds to rest between each trial. We collected 20 sets for each gesture in total. We created a tensor mapping window for each dataset over sensor data to the list of objects the user grasped. We combined all tensors and shuffled data for each subject to create a training dataset. We employed the hand joint angle estimation architecture we trained on earlier. We added two FC neural networks mapping hand joint angles to a list of objects. During the training phase, we updated the last two layers while freezing the rest of the network. More results can be found in Supplementary Data Figures 7, 8, and 9. More details about the properties of objects used in this study can be found in Supplementary Table 2. A demonstration video showcasing the object detection algorithm can be found in Supplementary Video 6.


\section*{Data Availability}
Data supporting this study's findings are available from project page, containing detailed explanations of all the datasets: \url{https://feel.ece.ubc.ca/SmartTextileGlove/}, as well as a direct link to a Google Drive Repository: \url{https://drive.google.com/drive/folders/1HWjG_6Y2G7XNEeI19Aids0g-dcufncGJ?usp=share_link} where the datasets can be downloaded.

\section*{Code Availability}
The codes supporting this study's findings are available from \url{https://github.com/arvintashakori/SmartTextileGlove}\cite{arvin_tashakori_2023_10128938}.

\section*{Acknowledgements}
The authors would like to thank support of NSERC-CIHR (CHRP549589-20, CPG-170611) awarded to PS, NSERC Discovery (NSERC: RGPIN-2017-04666 \& RGPAS-2017-507964) awarded to PS, NSERC Alliance (ALLRP 549207-19) awarded to PS, Mitacs (IT14342 \& IT11535) awarded to PS, and CFI and financial and technical support of Texavie Technologies Inc. and their staff.

\section*{Author Contributions Statement}

AT and PS developed system model and implemented the learning algorithm, iOS Mobile application, data pipeline, PC-based data acquisition software, Unity application, and firmware parts. ZJ and SS designed yarn-based strain sensors. ZJ, AS, SS, HN, KL, and PS developed hardware and fabricated gloves and sensors. AT performed the experiments and analysis with help and input from others. CN helped with PCB box fabrication and drawing sensor schematic. PS, AS, JJE, CY, and ZJW oversaw the project. All authors contributed to writing of the manuscript and analysis of results.

\section*{Competing interests}
PS, AT, ZJ, AS, SS and HN have filed a patent based on this work under the US provisional patent application no. 63/422,867. The remaining authors declare no competing interests.

~\newpage

\begin{savenotes}
\begin{table*}[ht]
\centering
\caption{\label{tab:compare}A comparison of overall performance parameters of the smart textile glove and ML system presented in this work and other related published works.}
\resizebox{1\textwidth}{!}{
\begin{tabular}{c c c c c c c c}
 & \shortstack{This Work} & \shortstack{Wen et al.\\(2021)}\cite{Wen2021} & \shortstack{Luo et al.\\(2021)}\cite{Luo2021} & \shortstack{Moin et al.\\(2021)}\cite{moin2021} & \shortstack{Zhou et al.\\(2020)}\cite{Zhou2020} & \shortstack{Hughes, et al.\\ (2020)}\cite{hughes2020simple}& \shortstack{Glauser et al.\\(2019)}\cite{Glauser2019}\\
\hline
 Form & Smart textile glove & Smart textile glove & Smart textile apparel & Flexible PCB armband & Smart textile glove & Smart textile gloves& Smart latex glove\\
 \hline

Sensors & \shortstack{25 HYSs +\\  2 $\times$9axis IMUs}  & \shortstack{15 triboelectric\\ sensors} &  \shortstack{$32\times32$ yarn-based piezo\\ resistive pressure sensors}& \shortstack{64 sEMG\\ sensors}&\shortstack{5 yarn-based \\strain sensors}&\shortstack{120 resistive knit sensors\\ + 6 fluidic pressure sensors}& \shortstack{44 Capacitive silicon\\ sensor array}\\
\hline

\multirow{7}{*}{Output} & fingers+wrist joint angles &  & \multirow{7}{*}{tactile feedback} & \multirow{7}{*}{21 gestures} & \multirow{7}{*}{11 gestures} &finger joint angle &\multirow{7}{*}{finger joint angles}\\
& keyboard typing & & & & & heart rate& \\
& 3D drawing & 50 dynamic gestures & & & &  10 handwritten letters&\\
& 50 dynamic gestures & + 20 sentences& & & & stiffness&\\
& 48 static gestures & & & & & temperature&\\
& 34 objects & & & & & 30 objects&\\
& force & & & & & force&\\
\hline

Robustness & \cmark &  \xmark & \cmark &  \xmark  &  \xmark  &  \xmark &  \xmark \\
\hline

Washability& \cmark & \xmark & \xmark &  \xmark  &  \xmark  &  \xmark &  \xmark \\
\hline

Wireless& \cmark &\xmark & \xmark&\cmark&\cmark&\xmark&\xmark\\
\hline

Sampling rate (Hz) & 20 & 600 & 14 & 1k&500&16&600\\
\hline

ML Response&\multirow{2}{*}{5} & \multirow{2}{*}{83} &\multirow{2}{*}{100}&\multirow{2}{*}{250}&\multirow{2}{*}{1000}&\multirow{2}{*}{Not Mentioned}&\multirow{2}{*}{125}\\
Time (ms) & & & & & & \\
\hline

Offline personalized& \multirow{2}{*}{1 minute} & \multirow{2}{*}{Not Mentioned} & \multirow{2}{*}{Not Mentioned} & \multirow{2}{*}{Not Mentioned} & \multirow{2}{*}{2 minutes} & MLP: $\sim$ 40 minutes& \multirow{1}{*}{2 hours}\\
training time& & & & & & LSTM: $\sim$ 3 hours&20 minutes, 1 minute\\
\hline

No GPU needed&\multirow{2}{*}{\cmark} & \multirow{2}{*}{\cmark} & \multirow{2}{*}{\xmark}& \multirow{2}{*}{\xmark}& \multirow{2}{*}{\xmark}& \multirow{2}{*}{\xmark}&\multirow{2}{*}{\xmark}\\
to calibrate& & & & & & \\
\hline

Finger joint angle & Intra: 1.24 deg&\multirow{2}{*}{\xmark}&\multirow{2}{*}{\xmark}&\multirow{2}{*}{\xmark}&\multirow{2}{*}{\xmark}&MLP: 6.40 deg&Intra: 5.8 deg (2h), 6.5 (20m)\\
estimation error & Inter: 1.45 deg& & & & &LSTM: 4.78 deg&Inter: 6.2 deg (2h), 7.6 (1m)\\
\hline

Static gesture & Intra: 97.81 \%&\multirow{2}{*}{Not Mentioned}&\multirow{2}{*}{\xmark}&Intra: 92.87 \%&3-fold cross-validation&\multirow{2}{*}{Not Mentioned}&\multirow{2}{*}{Not Mentioned}\\
recognition accuracy& Inter: 94.60 \% & & & Inter: 84.53 \% & 98.63 \%  & &\\
\hline

Dynamic gesture & Intra: 97.31 \%&50 words: 91.3 \%&\multirow{2}{*}{\xmark}&\multirow{2}{*}{\xmark}&\multirow{2}{*}{\xmark}&\multirow{2}{*}{Not Mentioned}&\multirow{2}{*}{Not Mentioned}\\
recognition accuracy& Inter: 94.05 \% & 20 Sentences: 95.0 \% & & &   & &\\
\hline

Typing accuracy & \shortstack{Intra: 97.8 \%}& Not Mentioned&\xmark&\xmark&\xmark&\xmark&\xmark\\
\hline

Object& Intra: 95.02 \%&\multirow{2}{*}{\xmark}&\multirow{2}{*}{\xmark}&\multirow{2}{*}{\xmark}&\multirow{2}{*}{\xmark}&MLP: 99.69 \% &\multirow{2}{*}{\xmark}\\
detection accuracy& Inter: 90.20 \% & & & &&LSTM: 98.18 \%&\\
\hline

Wrist angle&Intra: 2.09 deg&\multirow{2}{*}{\xmark}&\multirow{2}{*}{\xmark}&\multirow{2}{*}{\xmark}&\multirow{2}{*}{\xmark}&\multirow{2}{*}{\xmark}&\multirow{2}{*}{\xmark}\\
estimation error & Inter: 2.45 deg& & & &&\\
\hline\\
\multicolumn{8}{l}{Note: Please note that different advanced wearable systems reported in literature function based on different physiological information (e.g., EMG and tactile), sensing modalities and form factors and can}\\
\multicolumn{8}{l}{be used in variety of applications, which may not be fair to compare only based on performance parameters.}\\
\end{tabular}
}
\end{table*}
\end{savenotes}

~\newpage

\begin{figure*}[!htb]
     \centering
     \includegraphics[width=\textwidth]{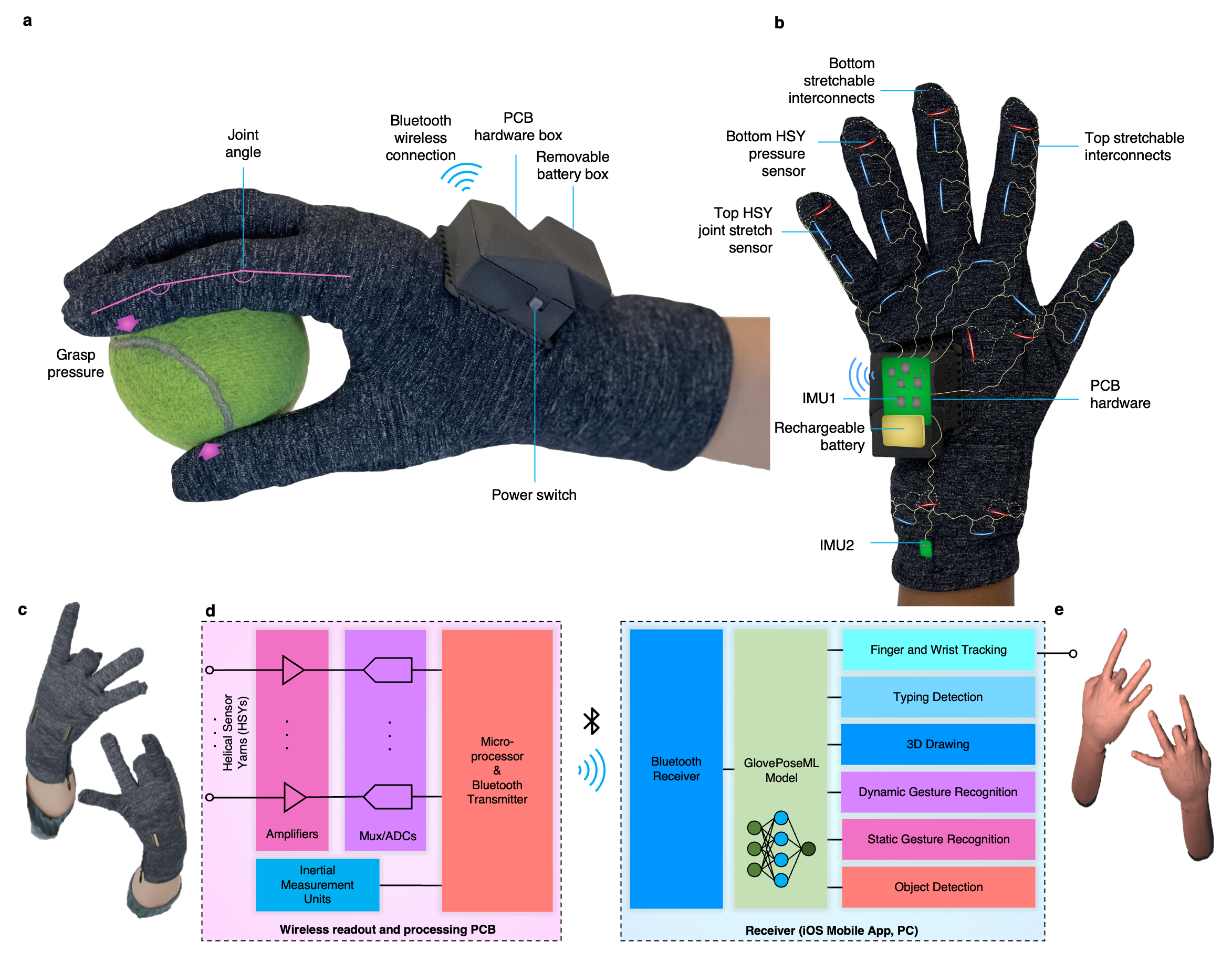}
\textbf{Fig. 1 Smart Textile Glove}. \textbf{a,} Photograph of the smart textile glove demonstrating its functionality in capturing joint angles and grasp pressure in interaction with objects. \textbf{b,} Photograph of the glove with X-ray schematic showing embedded HSYs (blue: top, red: bottom), 3D stretchy interconnects (gold lines, solid: top, dashed: bottom), the PCB including first IMU1 and other readout and Bluetooth hardware, battery box, and the second embedded IMU2, located just above the wrist. \textbf{c,} User wearing a glove pair showing a complex gesture. \textbf{d,} Schematic block diagram of how multiple HSYs and IMUs are connected to PCB hardware, including amplifiers, analog-to-digital-convertors (ADCs), microprocessor and Bluetooth low energy (BLE) transmitter and subsequently to iOS mobile app or a PC that receives the data and pass it to GlovePoseML model and demonstration apps for different applications. \textbf{e,} The visualization of user's complex hand gesture estimated by ML algorithm that dynamically follow the movements. \label{fig:sysmodel}
\end{figure*}

\newpage

\begin{figure*}[!htb]
     \centering
     \includegraphics[width=0.60\textwidth]{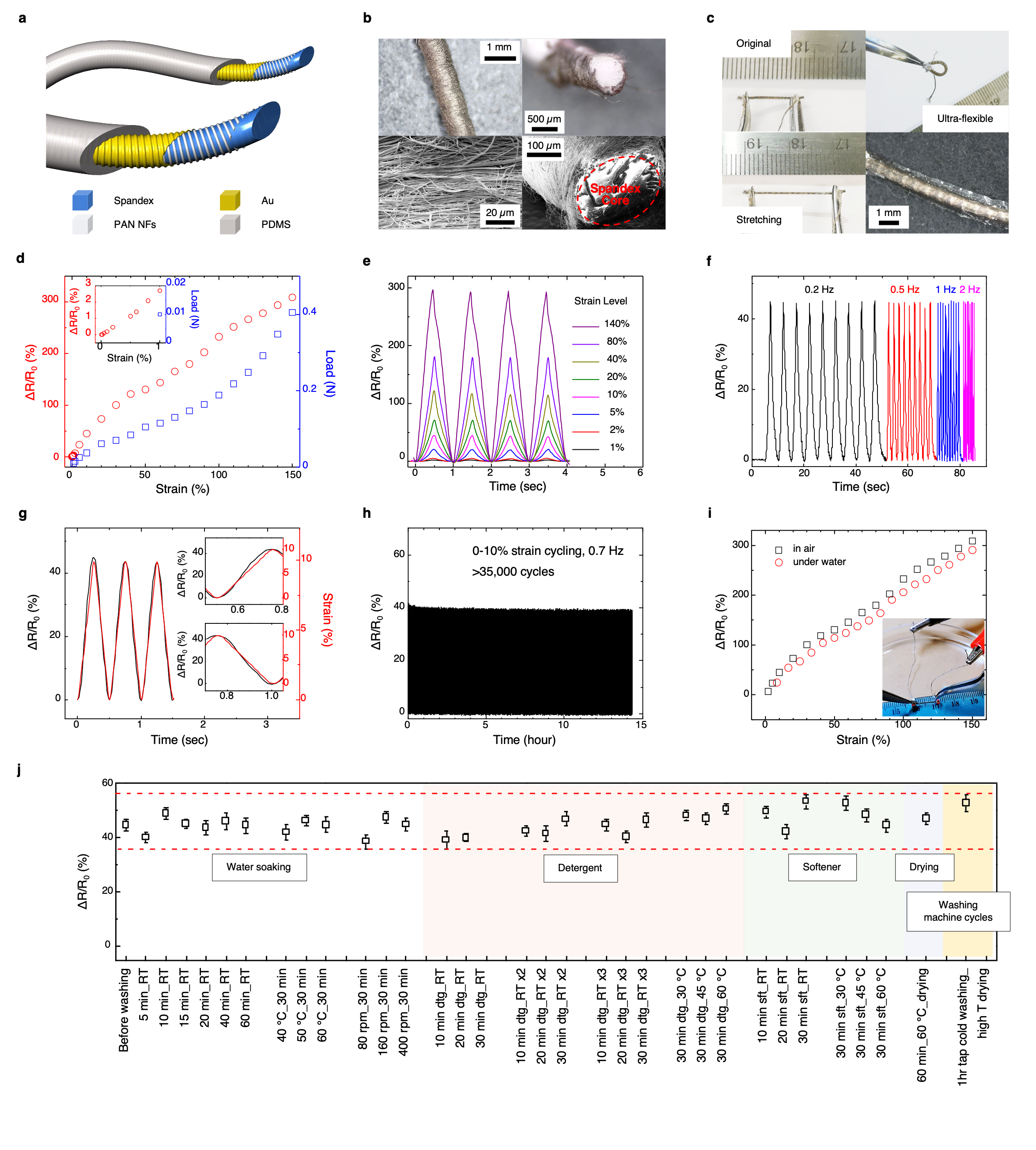}
     
     \textbf{Fig. 2 Helical Sensor Yarns. a,} Schematic of HSYs showing coaxial structure with elastic spandex core, wrapped with helical metal-coated NFs, and encapsulation elastomer shell. \textbf{b,} Microscope and SEM photomicrographs of HSYs before shell coating. \textbf{c,} Photographs of HSYs with shell and contact electrodes. \textbf{d,e} Measurement of sensitivity to different tensile strains and loads\textcolor{black}{, during loading and unloading, displaying exceptional sensitivity down to 0.005 \% strain and minimal hysteresis.} Inset: Sensitivity for strains < 1.0 \%. (The data points present the mean value of 20 samples. Error bars derived from standard deviation are too small to show on the figure.) \textbf{f,} Changes in resistance response of HSYs to strains up to a maximum of 10 \% strain with various frequencies 0.2 - 2 Hz. \textbf{g,} Sensor response accurately following changes in strain from 0 to 10 \%. Insets: Zoomed views of loading and unloading cycles. \textbf{h,} Mechanical durability test for up to 14 hours continuous stretch–release cycles.  \textbf{i,} Comparison of sensor response in air and underwater. Inset: Photograph of sensor during underwater tests. \textcolor{black}{(The data points present the mean value of 20 samples. Error bars derived from standard deviation are too low to show.)}  \textbf{j,} Durability sensor response in the smart textile glove during various laundry, washing and drying cycles.\label{fig:sensor}
\end{figure*}

\newpage

\begin{figure*}[!htb]
     \centering
     \includegraphics[width=0.7\textwidth]{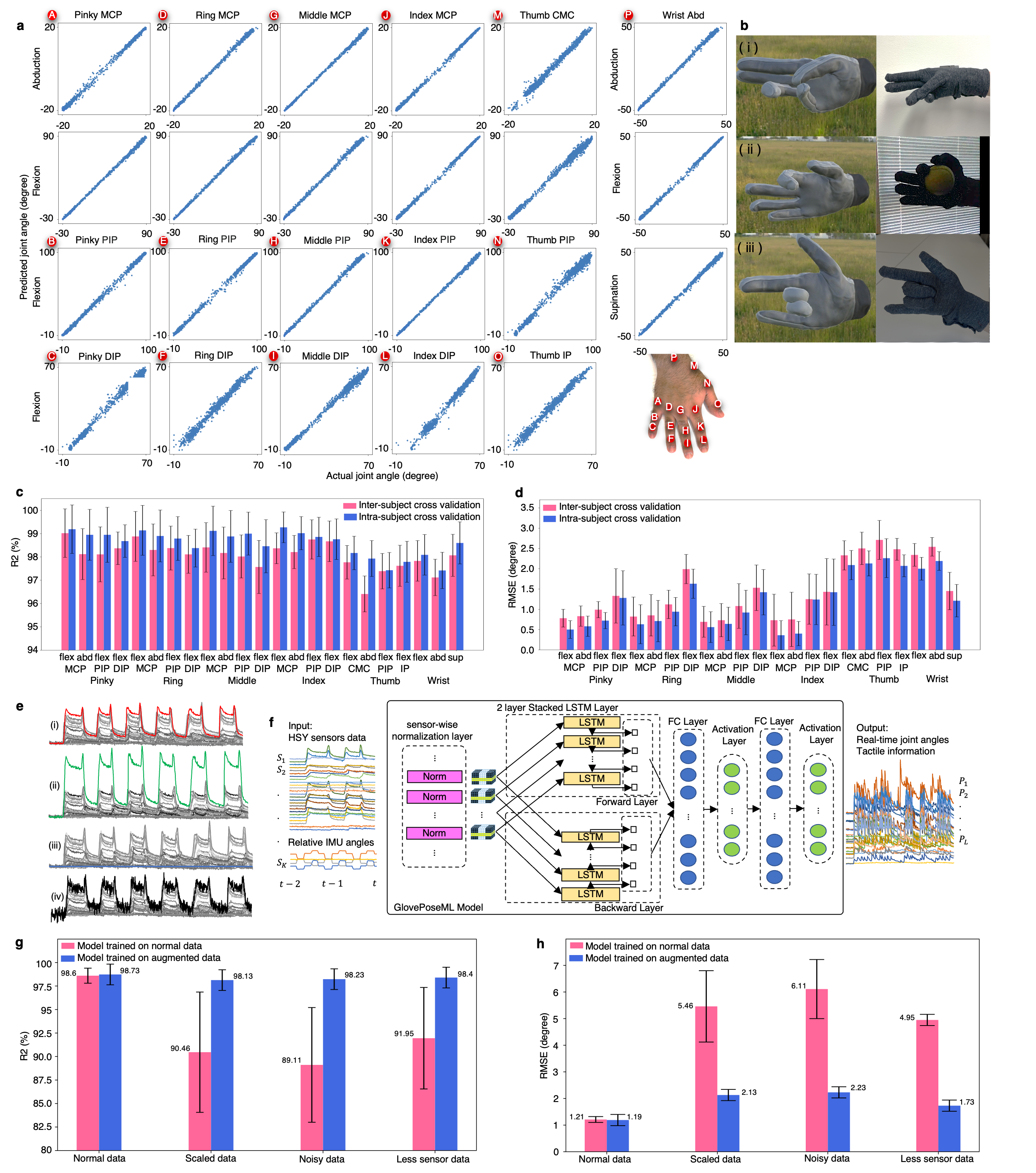}
     
     \textbf{Fig. 3 ML and Dynamic Hand Tracking. a,} Linear plot comparison of joint angles (A to P, shown in subset) estimated using GlovePoseML versus those measured by motion capture camera and marker system. \textbf{b,} Comparison of visualized estimated hand pose (left) and photographs (right) of complex movements in (i) normal conditions, (ii) when there is occlusion during grasping of a ball, and (iii) low light environment. Average accuracy results for different joints in terms of \textbf{c,} the goodness of fit R2, and \textbf{d,} RMSE. \textbf{e,} Scenarios for data-augmentation: (i) original data, (ii) scaled data, (iii) data with fewer active sensors, and (iv) noisy data. \textbf{f,} Overall architecture of tracking GlovePoseML model, showing the normalization layer, 2-layer stacked Bi-LSTM model, 2 FC layers and activation layers. Comparison of average accuracy results of model trained using normal and augmented dataset in terms of \textbf{g,} goodness of fit R2, and \textbf{h,} RMSE.
      \label{fig:handpose}
\end{figure*}

\newpage

\begin{figure*}[!htb]
     \centering
     \includegraphics[width=0.7\textwidth]{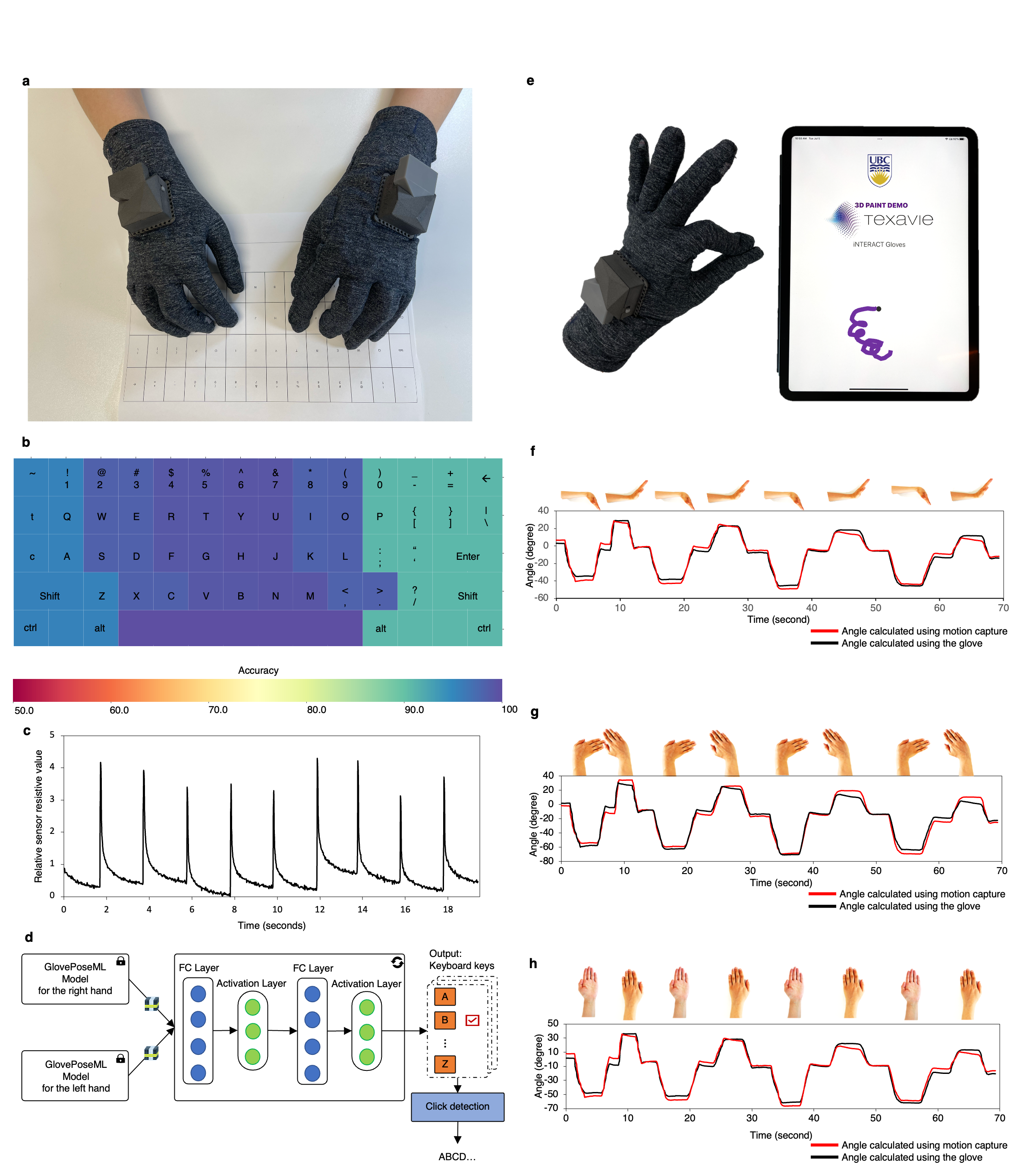}
     
     \textbf{Fig. 4 Typing and Drawing Demos. a, } Photograph of user using a pair of wireless smart gloves for typing on a mock paper keyboard used for user's visual feedback. \textbf{b,} Color-coded comparison of accuracy of detection of each key on the keyboard in a 10-finger typing on the mock paper keyboard. \textbf{c,} Typical response of HSY sensor at the tip of a finger as it repeatedly taps and retracts the surface of mock paper keyboard. \textbf{d,} schematic of the two FC and activation layers that work with GlovePoseML model for detection of typing. \textbf{e,} Illustration of a 3D in-air drawing application on an iPad based on two-finger pinch and wrist movement. Comparison between the estimated angle of the wrist using our smart glove and ML system and gold standard motion capture system for tracking \textbf{f,} wrist flexion and extension, \textbf{g,} wrist abduction and adduction, and \textbf{h,} wrist supination and pronation.
      \label{fig:application1}
\end{figure*}

~\newpage

\begin{figure*}[!ht]
     \centering
     \includegraphics[width=0.7\textwidth]{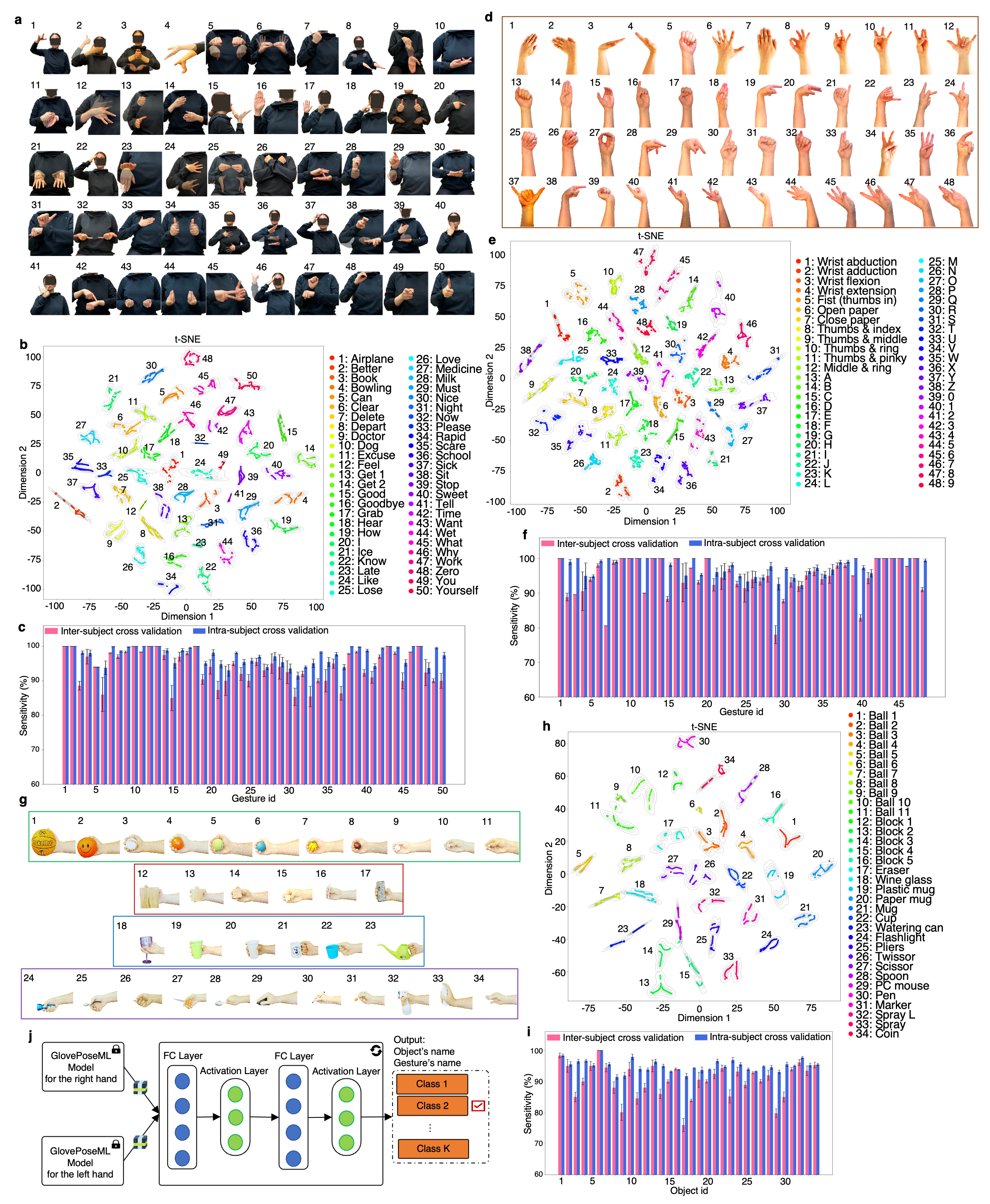}
     
     \textbf{Fig. 5 Real-time dynamic/static gesture, and object recognition. a,} Pictures of dynamic gestures used for hand gesture recognition. \textbf{b,} Cluster distribution of dynamic gestures from GlovePoseML output layer between users. \textbf{c,} Sensitivity (\%) results for dynamic gestures in inter- and intra-subject cross-validation. \textbf{d,} Pictures of static gestures used for hand gesture recognition. \textbf{e,} Cluster distribution of static gestures from GlovePoseML output layer between users. \textbf{f,} Sensitivity (\%) results for static gestures in inter- and intra-subject cross-validation. \textbf{g,} Pictures of objects used for object recognition from grasp form study. \textbf{h,} Cluster distribution of objects from GlovePoseML output layer between users. \textbf{i,} Sensitivity (\%) results for different objects in inter- and intra-subject cross-validation. \textbf{j,} Schematic of the model used for gesture and object classification, displaying FC and activation layers.
      \label{fig:objectandgestures}
\end{figure*}

\newpage
\begin{figure*}[!ht]
  \centering
  \includegraphics[width=0.7\linewidth]{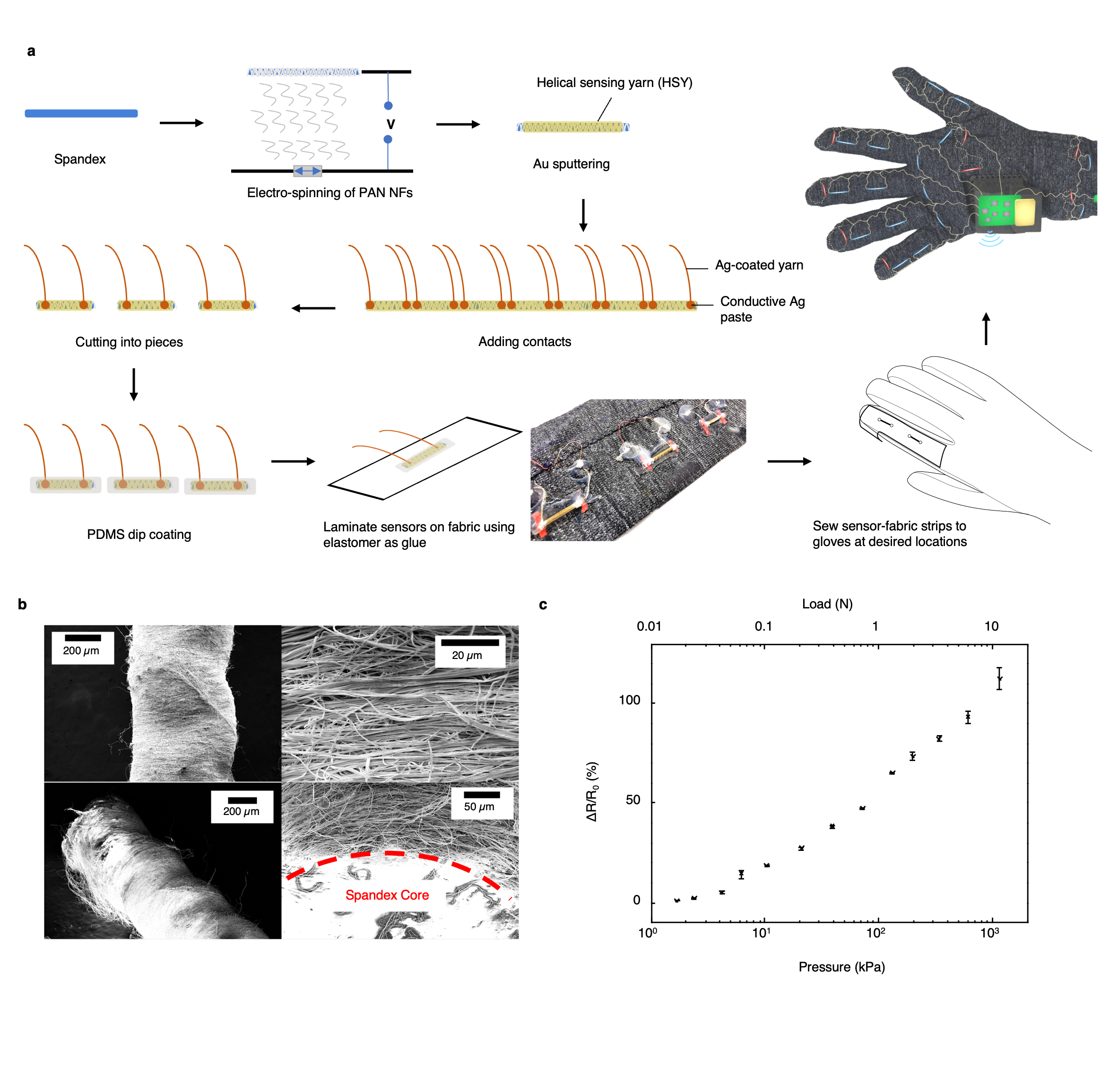}
  \label{fig:Sensor_S1}
  
  \textbf{Extended Data Figure 1. Fabrication and Characteristics of HSYs.} \textbf{a,} Fabrication process of HSYs and gloves. \textbf{b,} SEM images of the yarn sensors before PDMS coating. \textbf{c,} Sensitivity of HSY resistance to various compressive pressure values applied at a frequency of 1 Hz. A metal indenter with the size of $5 mm \times 5 mm$ were used to apply pressure normal to the fabric. (The data points present the mean values $\pm$ standard deviation of 20 samples.) \textbf{d,} The strain sensitivity of our insulated yarn sensors made from optimized composites of carbon particles with highly stretchable elastomers, demonstrating high stretchability up to 1,000 \%, but showing more hysteresis for > 500 \% stretch and slower responsiveness due to the softer nature of these sensors in comparison to HSYs. This highlights the superior performance of HSYs for the proposed smart glove real-time applications with less than 120 \% maximum stretch for the fabric.
\end{figure*}

\newpage

\begin{figure}[h]
  \centering
  \includegraphics[width=1\linewidth]{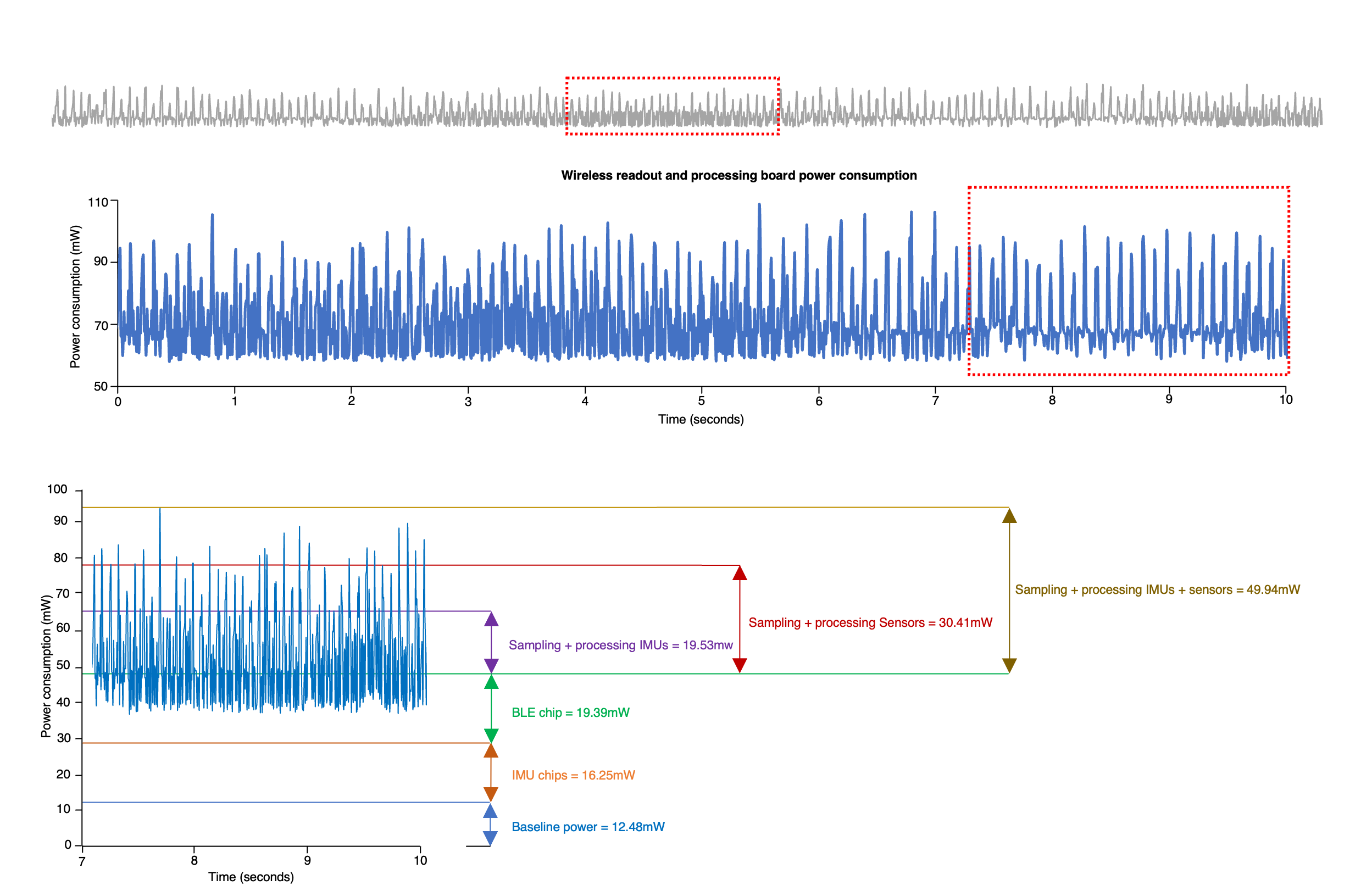}
  \label{fig:power}

\textbf{Extended Data Figure 2. Power Consumption} Custom-made wireless board power consumption breakdown for different components including BLE chip, IMU chips, and all HSYs.
\label{fig:bordpower}
\end{figure}

\newpage

\begin{figure*}[!ht]
     \centering
     \includegraphics[width=0.8\textwidth]{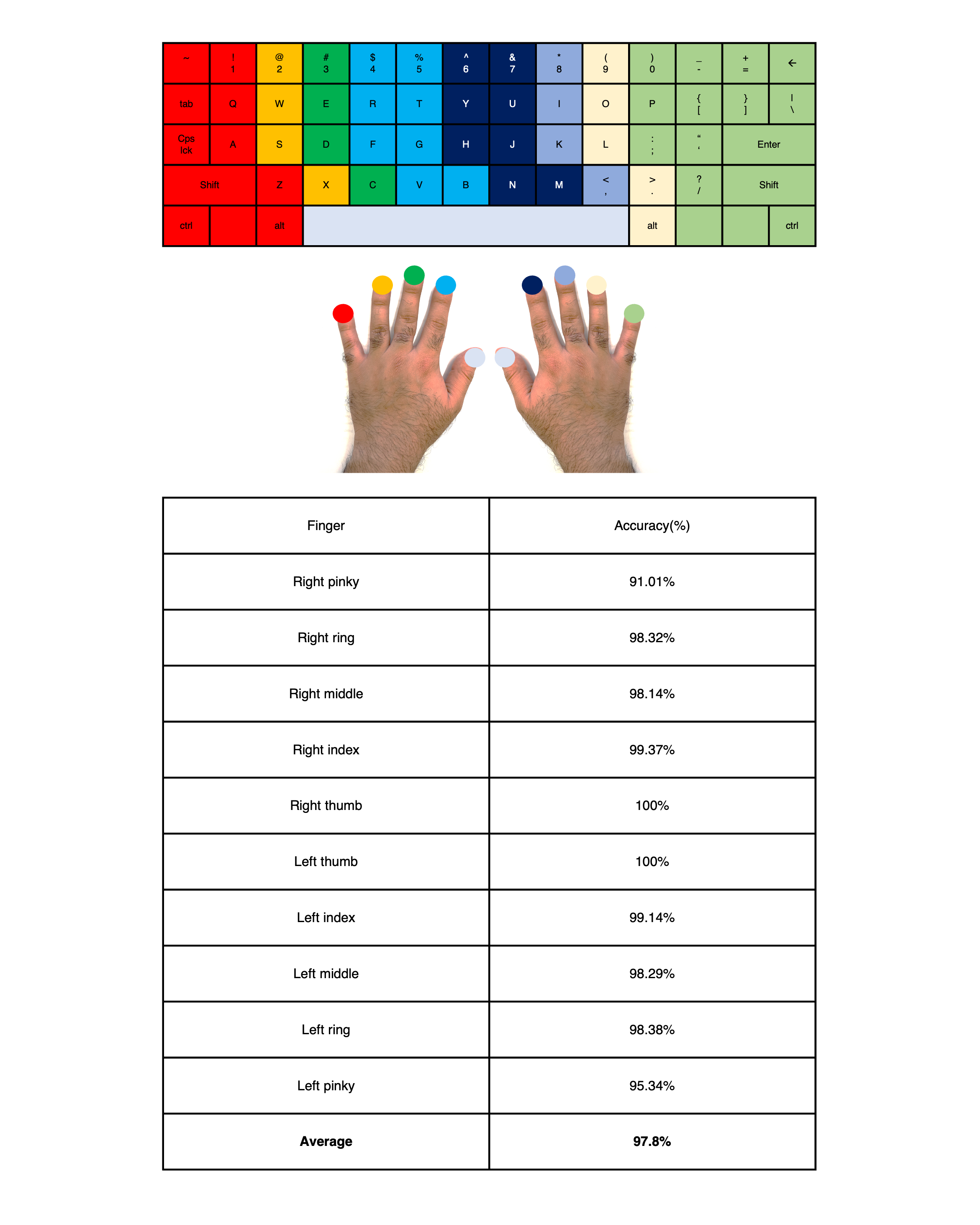} \label{fig:Keyboardextended1}
     
     \textbf{Extended Data Figure 3. Keyboard typing detection} Inter-session cross-validation accuracy results for the keyboard typing detection algorithm.
\end{figure*}

\newpage

\begin{figure*}[h]
     \centering
     \includegraphics[width=0.7\textwidth]{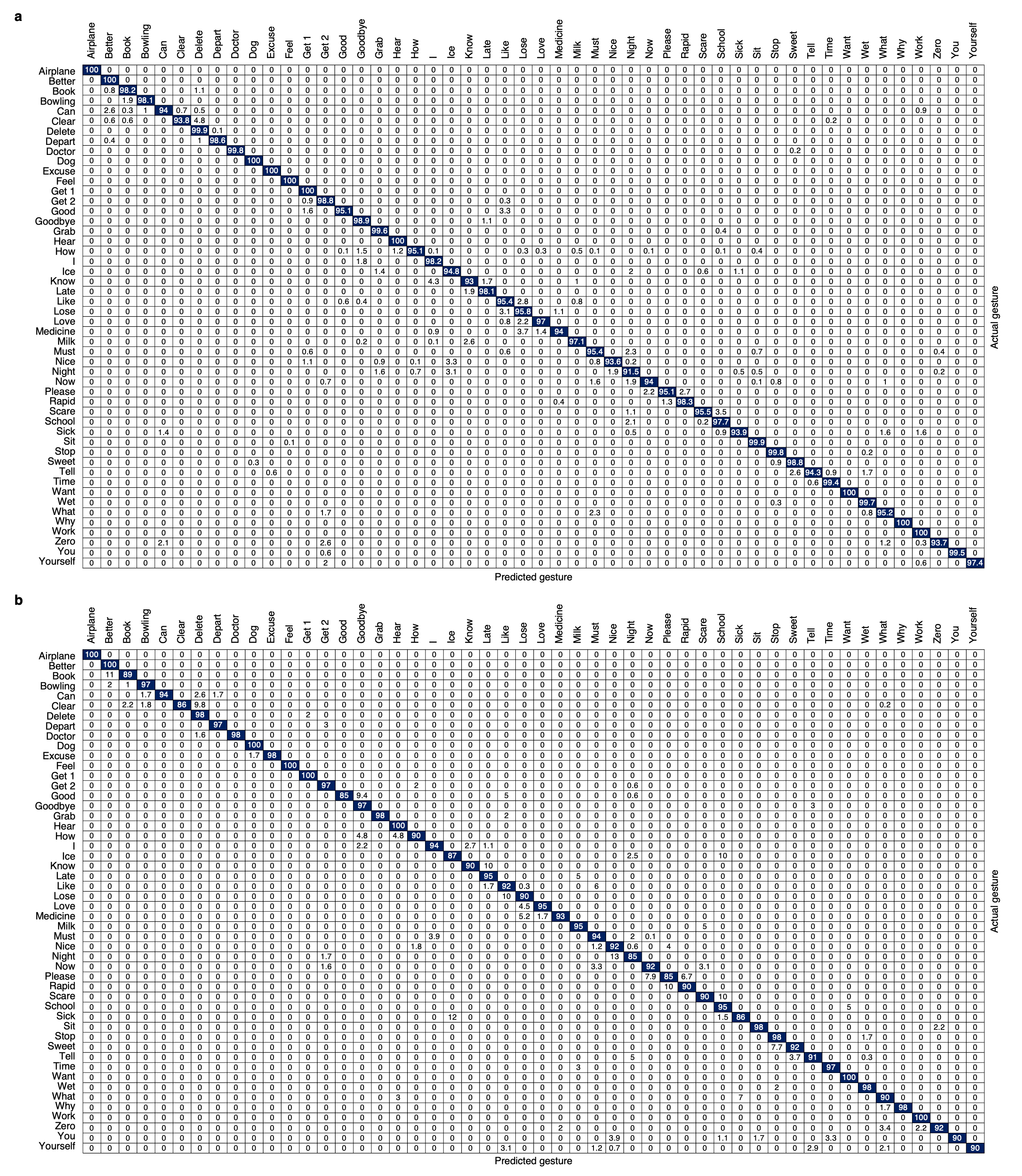}
     \label{fig:dynamicgestureextended}
     
     \textbf{Extended Data Figure 4. Dynamic gesture recognition} Confusion matrix for \textbf{a,} Intra-subject (accuracy: 97.31 \%), and \textbf{b,} Inter-subject cross-validation (accuracy: 94.05 \%).
\end{figure*}

\newpage

\begin{figure*}[h]
     \centering
     \includegraphics[width=0.7\textwidth]{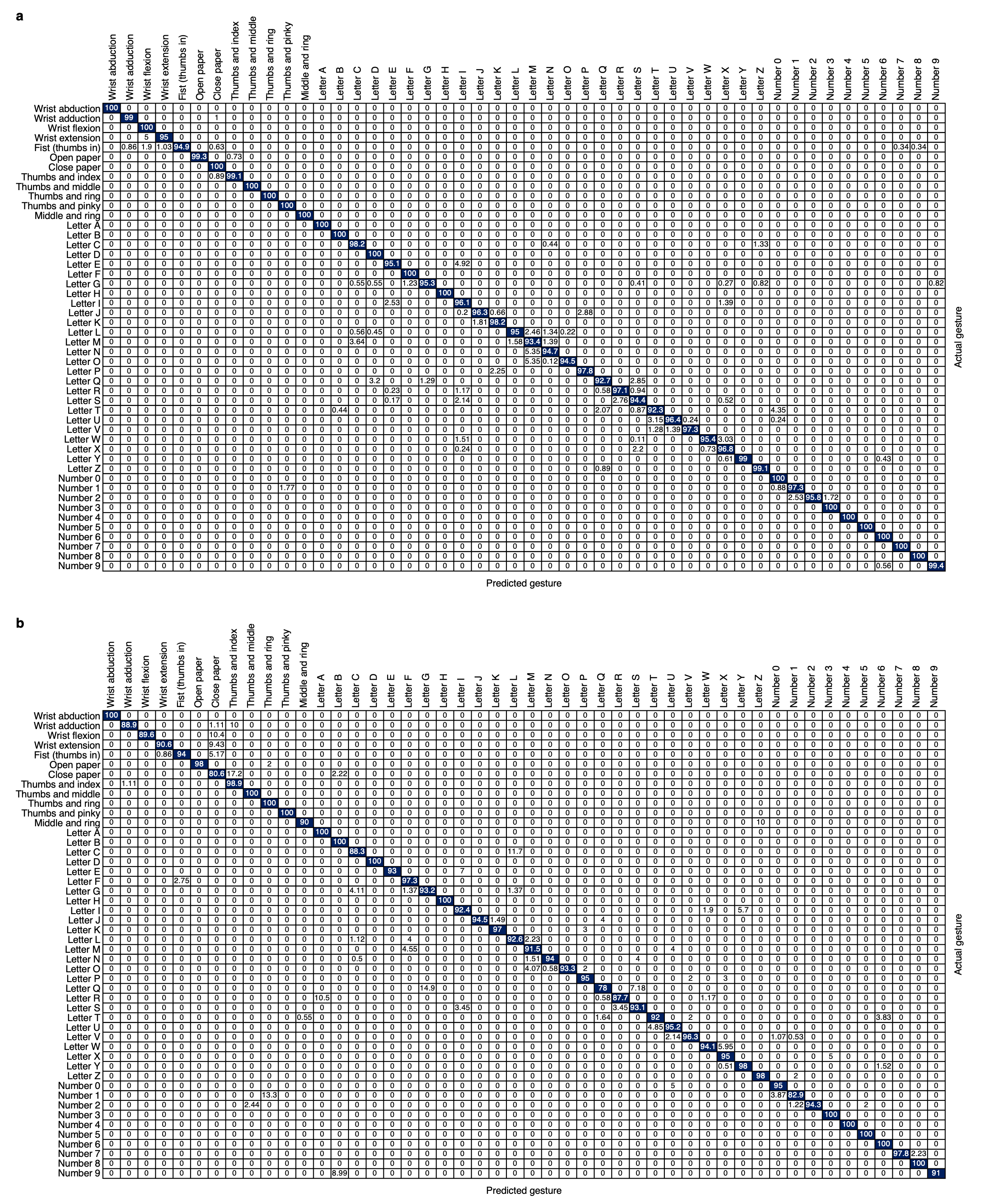}
     \label{fig:staticgestureextended}
     
     \textbf{Extended Data Figure 5. Static gesture recognition} Confusion matrix for \textbf{a,} Intra-subject (accuracy: 97.81 \%), and \textbf{b,} Inter-subject cross-validation (accuracy: 94.60 \%).
\end{figure*}

\newpage

\begin{figure*}[h]
     \centering
     \includegraphics[width=0.9\textwidth]{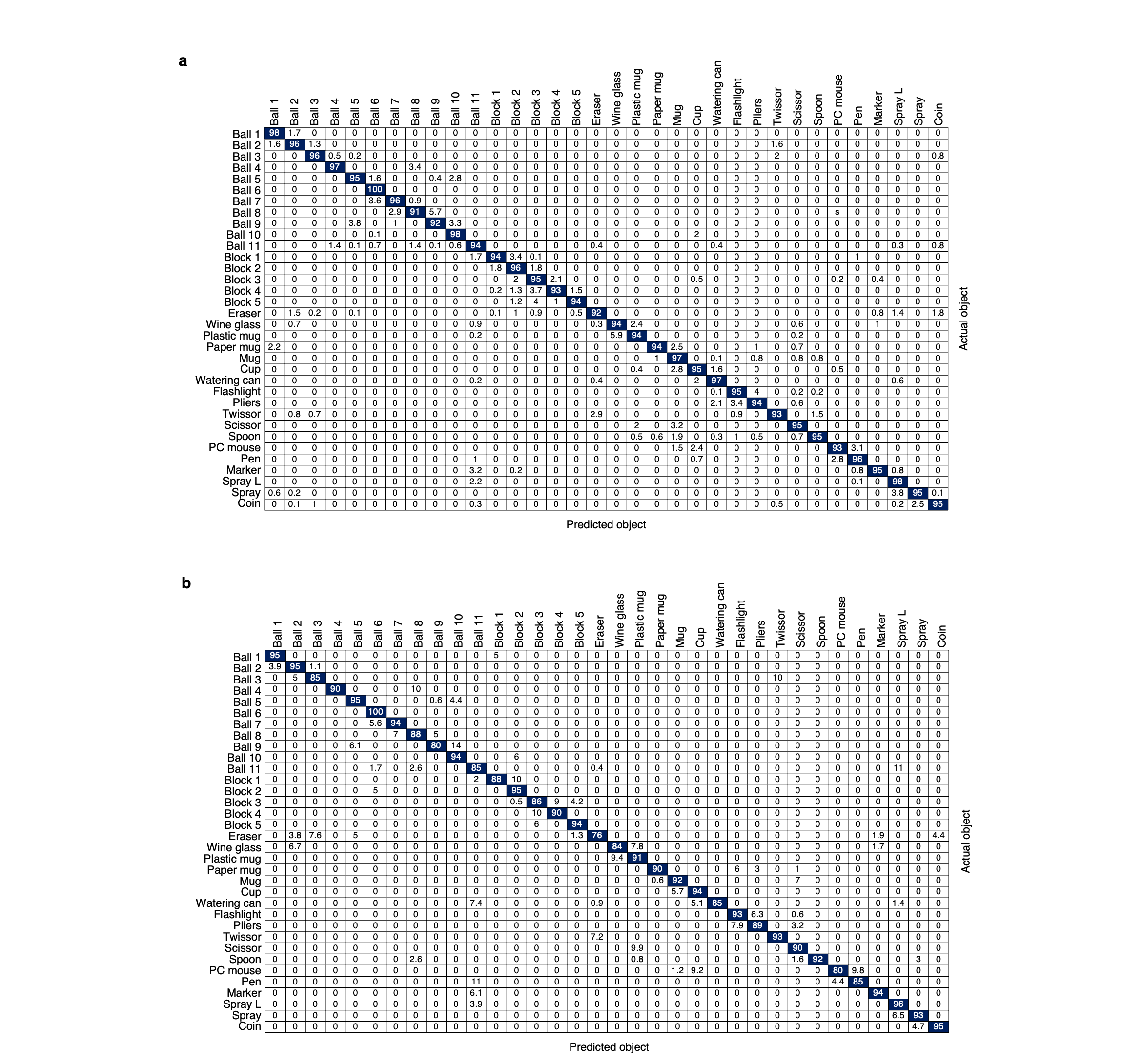}
     \label{fig:objectextended}
     
     \textbf{Extended Data Figure 6. Object detection} Confusion matrix for \textbf{a,} Intra-subject (accuracy: 95.02 \%), and \textbf{b,} Inter-subject cross-validation (accuracy: 90.20 \%).
\end{figure*}

 \newpage

 \section*{Supplementary Algorithm 1. Data Augmentation Algorithm}\label{selfsupp_app}

\begin{algorithm}
\caption{Data Augmentation Algorithm}\label{alg:cap}
\begin{algorithmic}
\Require \text{Labeled dataset D, Transformation functions T, Number of  epochs E}
\Ensure \text{Trained multitask model M}
\State $\text{D}_{\text{aug}} \gets [ ]$
\For{\text{(signals, label) in D}}
    \State $\text{Y} \gets [0,0,...,0]$
    \State $\text{D}_{\text{aug}}.\text{append}(\text{signals}, \text{label},\text{Y})$
    \For{\text{Transformation in T}}
    \State $\text{Y[Transformation]=1}$
    \State $\text{D}_{\text{aug}}.\text{append}(\text{Transformation(signals)}, \text{label},\text{Y})$
    \State $\text{Y[Transformation]=0}$
    \EndFor
\EndFor
\State $\text{M} \gets \text{New Multitask model}$
\For{\text{Epoch in E}}
\State $\text{TrainingMinibatch(model = M, dataset = D}_{\text{aug}}\text{, loss = L1($\beta = 0.5$) +}$ \par
        \hskip\algorithmicindent \text{BinaryTransformationLoss (), optimizer = Adam($lr=0.0001$))}
\EndFor
\end{algorithmic}
\end{algorithm}

\begin{landscape}
\section*{Supplementary Table 1. Hand joint angle estimation}\label{secA5}
\begin{table}[h]
\centering
\label{table:handposeresults}
\caption{Detailed inter-subject and intra-subject validation results for finger and wrist joint angles.}
\resizebox{0.8\paperheight}{!}{
\begin{tabular}{|c|c|c|c|c|c|c|c|c|c|c|c|c|c|c|c|c|c|c|c|c|c|c|c|c|} 
\hline
\multirow{3}{*}{Subject}          & \multirow{3}{*}{Performance} & \multicolumn{4}{c}{Pinky}              & \multicolumn{4}{c}{Ring}              & \multicolumn{4}{c}{Middle}            & \multicolumn{4}{c}{Index}             & \multicolumn{3}{c}{Thumb}      & \multicolumn{3}{c}{Wrist}                                          & \multirow{3}{*}{Average}  \\
                                  &                              & \multicolumn{2}{c}{MCP} & PIP   & DIP  & \multicolumn{2}{c}{MCP} & PIP  & DIP  & \multicolumn{2}{c}{MCP} & PIP  & DIP  & \multicolumn{2}{c}{MCP} & PIP  & DIP  & \multicolumn{2}{c}{MCP} & IP   & \multirow{2}{*}{Flex} & \multirow{2}{*}{Abd} & \multirow{2}{*}{Sup} &                           \\
                                  &                              & Flex  & Abd              & Flex  & Flex & Flex  & Abd              & Flex & Flex & Flex  & Abd              & Flex & Flex & Flex  & Abd              & Flex & Flex & Flex  & Abd              & Flex &                       &                      &                      &                           \\ 
\hline
\multirow{2}{*}{Subject 1}        & RMSE                         & 0.34  & 0.38             & 0.59  & 0.87 & 0.38  & 0.42             & 0.79 & 1.39 & 0.41  & 0.45             & 0.62 & 1.09 & 0.19  & 0.23             & 0.87 & 1.03 & 1.83  & 1.87             & 2.02 & 1.93                  & 1.87                 & 2.12                 & 0.99                      \\
                                  & R2                           & 99.85 & 99.75            & 99.56 & 99   & 99.73 & 99.63            & 99.3 & 98.9 & 99.69 & 99.6             & 99.5 & 99   & 99.54 & 99.4             & 99.2 & 99.1 & 98.62 & 98.5             & 97.6 & 98.2                  & 98                   & 97.6                 & 99.07                     \\
\multirow{2}{*}{Subject 2}        & RMSE                         & 0.41  & 0.47             & 0.65  & 1.01 & 0.37  & 0.43             & 0.7  & 1.43 & 0.35  & 0.41             & 0.68 & 1.2  & 0.2   & 0.26             & 1.01 & 1.01 & 1.94  & 2                & 1.79 & 1.84                  & 1.73                 & 2.03                 & 1                         \\
                                  & R2                           & 99.55 & 99.21            & 99.52 & 99   & 99.7  & 99.36            & 99.2 & 98.7 & 99.45 & 99.1             & 99.4 & 99   & 99.61 & 99.3             & 99.3 & 99.3 & 98.5  & 98.2             & 98.1 & 98.3                  & 99.1                 & 97.9                 & 99.04                     \\
\multirow{2}{*}{Subject 3}        & RMSE                         & 0.89  & 1.02             & 1.07  & 2.47 & 1.49  & 1.62             & 1.56 & 2.25 & 1.24  & 1.37             & 1.9  & 2.41 & 1.01  & 0.94             & 2.34 & 2.89 & 2.73  & 2.66             & 3.02 & 2.53                  & 2.46                 & 2.57                 & 1.93                      \\
                                  & R2                           & 97.35 & 97.04            & 96.84 & 97.4 & 97.25 & 96.94            & 97.1 & 96.9 & 97.23 & 96.9             & 97.4 & 96.4 & 98.08 & 97.8             & 97.4 & 97.2 & 96.89 & 96.6             & 96.1 & 96.2                  & 96.7                 & 96.1                 & 96.99                     \\
\multirow{2}{*}{Subject 4}        & RMSE                         & 0.45  & 0.52             & 0.67  & 1.07 & 0.49  & 0.56             & 0.86 & 1.62 & 0.42  & 0.49             & 0.72 & 1.21 & 0.21  & 0.28             & 1.01 & 1.12 & 2.01  & 2.08             & 2.39 & 1.95                  & 1.89                 & 2.01                 & 1.09                      \\
                                  & R2                           & 99.55 & 99.43            & 99.34 & 98.9 & 99.42 & 99.3             & 99.1 & 98.6 & 99.63 & 99.5             & 99.4 & 98.9 & 99.54 & 99.4             & 99.1 & 99   & 98.34 & 98.2             & 97.6 & 98                    & 98.2                 & 97.6                 & 98.92                     \\
\multirow{2}{*}{Subject 5}        & RMSE                         & 0.4   & 0.49             & 0.64  & 0.98 & 0.41  & 0.5              & 0.78 & 1.48 & 0.39  & 0.48             & 0.67 & 1.17 & 0.2   & 0.29             & 0.96 & 1.05 & 1.93  & 2.02             & 2.07 & 2.12                  & 2.06                 & 2.2                  & 1.06                      \\
                                  & R2                           & 99.65 & 99.31            & 99.47 & 99   & 99.62 & 99.28            & 99.2 & 98.7 & 99.59 & 99.3             & 99.4 & 99   & 99.56 & 99.2             & 99.2 & 99.1 & 98.49 & 98.2             & 97.8 & 98.2                  & 98.4                 & 98                   & 98.98                     \\ 
\hline
\multirow{2}{*}{Intra subject CV} & RMSE                         & 0.5   & 0.58             & 0.72  & 1.28 & 0.63  & 0.71             & 0.94 & 1.63 & 0.56  & 0.64             & 0.92 & 1.42 & 0.36  & 0.4              & 1.24 & 1.42 & 2.09  & 2.13             & 2.26 & 2.07                  & 2                    & 2.19                 & 1.21                      \\
                                  & R2                           & 99.19 & 98.95            & 98.95 & 98.7 & 99.14 & 98.9             & 98.8 & 98.4 & 99.12 & 98.9             & 99   & 98.5 & 99.27 & 99               & 98.9 & 98.8 & 98.17 & 97.9             & 97.4 & 97.8                  & 98.1                 & 97.4                 & 98.6                      \\ 
\hline
\multirow{2}{*}{Inter subject CV} & RMSE                         & 0.78  & 0.83             & 0.99  & 1.33 & 0.82  & 0.85             & 1.12 & 1.99 & 0.69  & 0.73             & 1.08 & 1.53 & 0.73  & 0.75             & 1.25 & 1.43 & 2.33  & 2.5              & 2.71 & 2.48                  & 2.34                 & 2.54                 & 1.45                      \\
                                  & R2                           & 99.02 & 98.12            & 98.11 & 98.4 & 98.88 & 98.3             & 98.4 & 98.1 & 98.41 & 98.2             & 98   & 97.6 & 98.37 & 98.2             & 98.8 & 98.7 & 97.77 & 96.4             & 97.4 & 97.6                  & 97.8                 & 97.1                 & 98.07                     \\
\hline
\end{tabular}
}
\end{table}
\end{landscape}

\section*{Supplementary Algorithm 2. Tap and touch detection algorithm}\label{click}

The objective of the tap detection algorithm is to find:

\vspace{-25pt}

\begin{equation}
T(t)=[T_{1}(t), T_{2}(t),...,T_{10}(t)]
\end{equation}

\vspace{-15pt}

where each 
$T_{i}(t)$
is a binary function representing the status of taps detected on the tip of i-th finger. We define a cost function measuring the sensor changes and comparing it with a predefined threshold:

\vspace{-15pt}

\begin{equation}
f_{i}(t) =
  \begin{cases}
    1      & \quad \text{if } \Big(\frac{S_{i}(t)}{S_{i}(0)}-1\Big)^2 \geq \text{Threshold}\\
    0  & \quad \text{if } \Big(\frac{S_{i}(t)}{S_{i}(0)}-1\Big)^2 < \text{Threshold}
  \end{cases}
\end{equation}

\vspace{-15pt}

Where $S_{i}(t)$ is the resistive value of sensor $i$ at the time $t$, and $S_{i}(0)$ is the initial resistive value of sensor i for the rest time. clicks are the number of rising edges on $f_{i}(t)$. To make the algorithm more stable, we look into the last four received samples and define our rising edge detection function as below:

\vspace{-25pt}

\begin{equation}
T_{i}(t) =
  \begin{cases}
    1      & \quad \text{if } f_{i}(t-3)=0 \& f_{i}(t-2)=0 \& f_{i}(t-1) = 1 \& f_{i}(t)=1\\
    0  & \quad \text{Otherwise}
  \end{cases}
\end{equation}

\vspace{-15pt}

For the drawing app, the user can select different colors by touching the thumb to the rest of the fingers. We employ the cost function we developed for the tap detection algorithm earlier. The color selection protocol is as below:

\vspace{-15pt}

\begin{itemize}
    \item $f_{Thumb} \times f_{Index}=1 \to \text{purple selected}$
    \item $f_{Thumb}\times f_{Middle}=1 \to \text{red selected}$
    \item $f_{Thumb}\times f_{Ring}=1 \to \text{blue selected}$
    \item $f_{Thumb}\times f_{Pinky}=1 \to \text{green selected}$
\end{itemize}

\newpage

\section*{Supplementary Data Figure 1. Detailed dynamic gesture recognition - Part 1}

\begin{figure*}[h]
     \centering
     \includegraphics[width=0.7\textwidth]{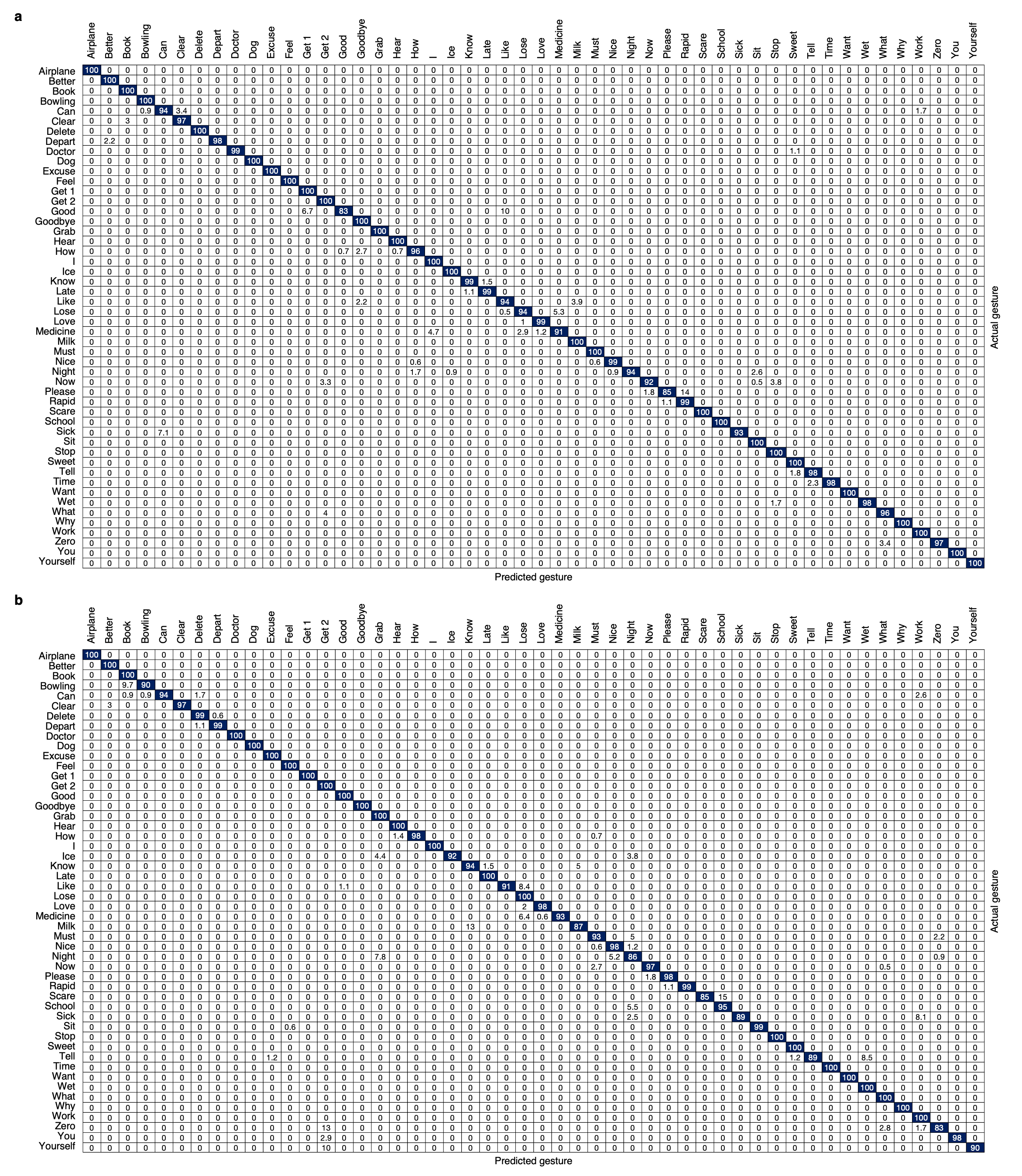}
     \label{fig:supp_1}
     \caption{Inter-session cross-validation confusion matrix for \textbf{a,} subject 1 (accuracy: 97.82 \%), and \textbf{b,} subject 2 (accuracy: 96.61 \%).}
      
\end{figure*}

\newpage

\section*{Supplementary Data Figure 2. Detailed dynamic gesture recognition - Part 2}

\begin{figure*}[h]
     \centering
     \includegraphics[width=0.7\textwidth]{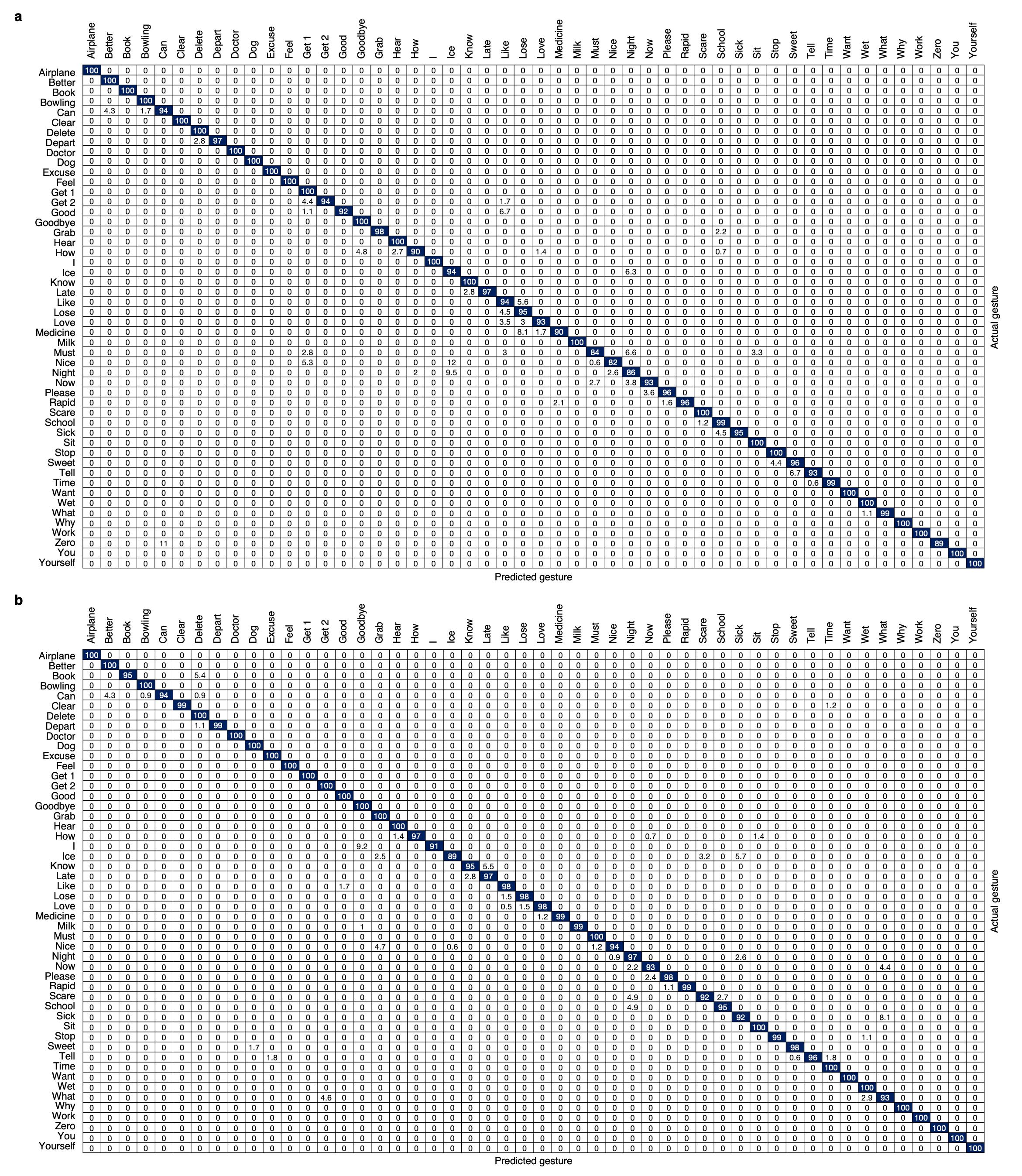}
     \label{fig:supp_2}
     \caption{Inter-session cross-validation confusion matrix for \textbf{a,} subject 3 (accuracy: 96.78 \%), and \textbf{b,} subject 4 (accuracy: 97.83 \%).}
      
\end{figure*}

\newpage

\section*{Supplementary Data Figure 3. Detailed dynamic gesture recognition - Part 3}

\begin{figure*}[h]
     \centering
     \includegraphics[width=0.7\textwidth]{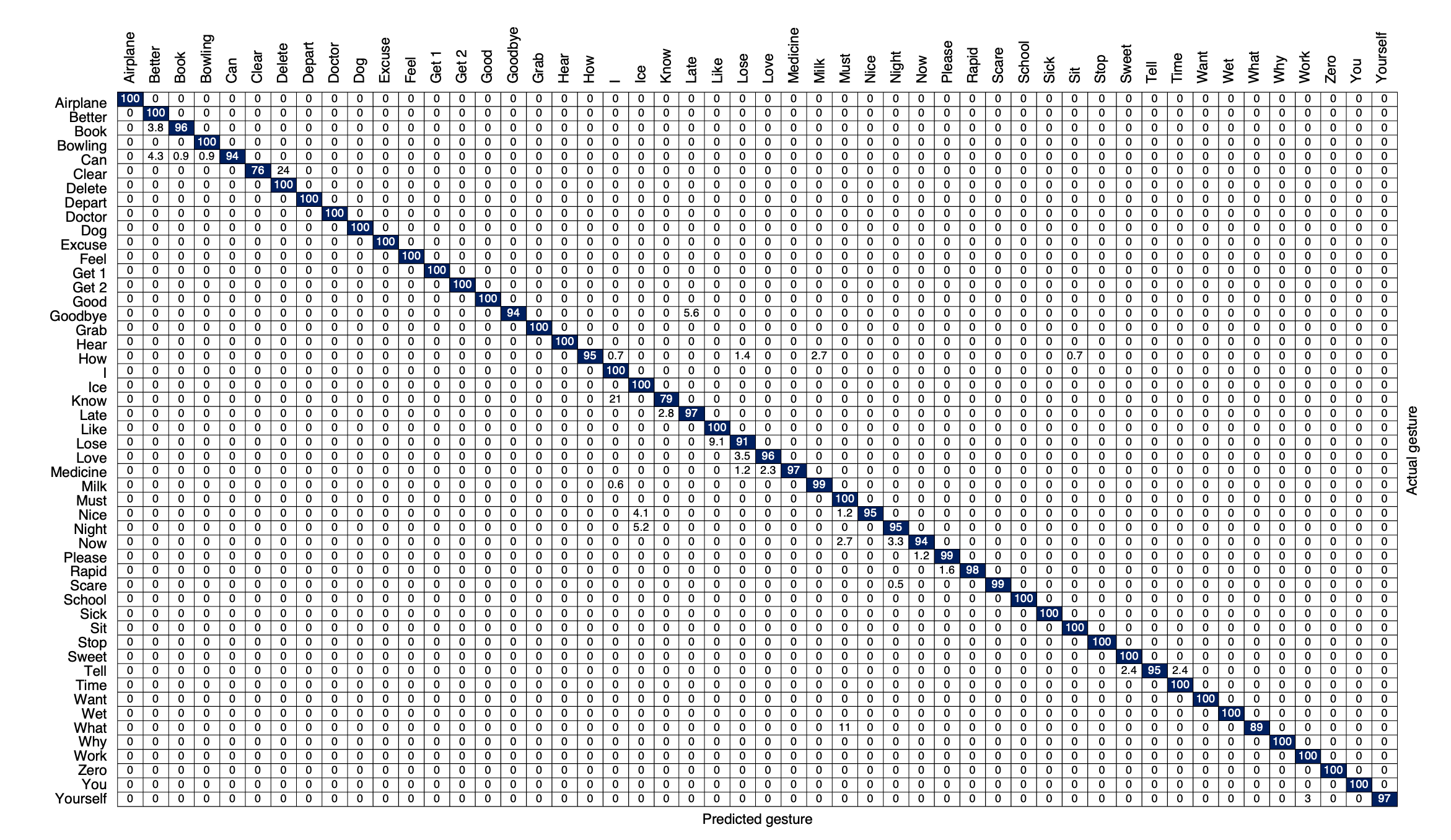}
     \label{fig:supp_3}
     \caption{Inter-session cross-validation confusion matrix for subject 5 (accuracy: 97.51 \%).}
      
\end{figure*}

\newpage

\section*{Supplementary Data Figure 4. Detailed static gesture recognition - Part 1}

\begin{figure*}[h]
     \centering
     \includegraphics[width=0.7\textwidth]{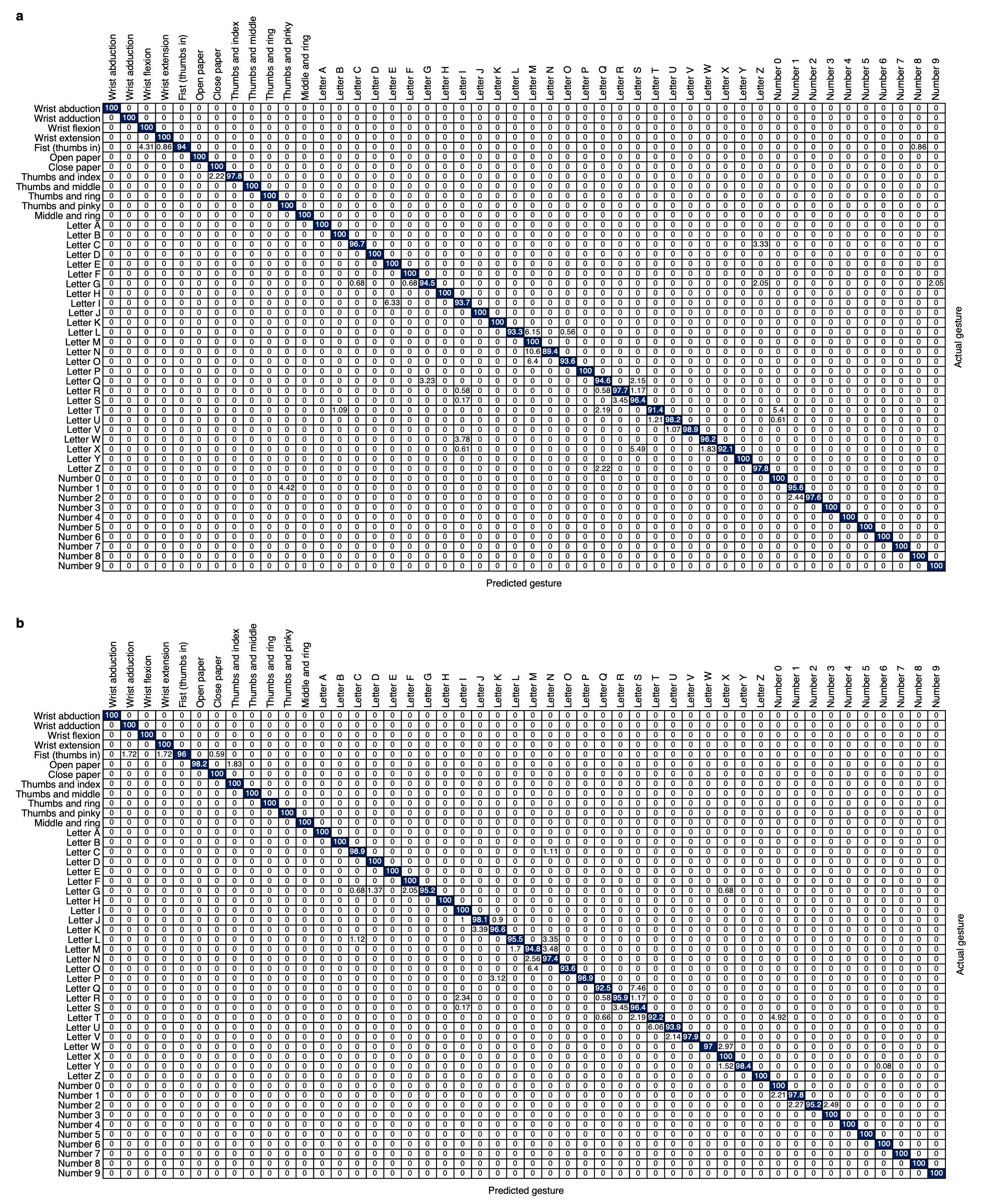}
     \label{fig:supp_4}
     \caption{Inter-session cross-validation confusion matrix for \textbf{a,} subject 1 (accuracy: 98.11 \%), and \textbf{b,} subject 2 (accuracy: 98.30 \%).}
      
\end{figure*}

\section*{Supplementary Data Figure 5. Detailed static gesture recognition - Part 2}

\begin{figure*}[h]
     \centering
     \includegraphics[width=0.7\textwidth]{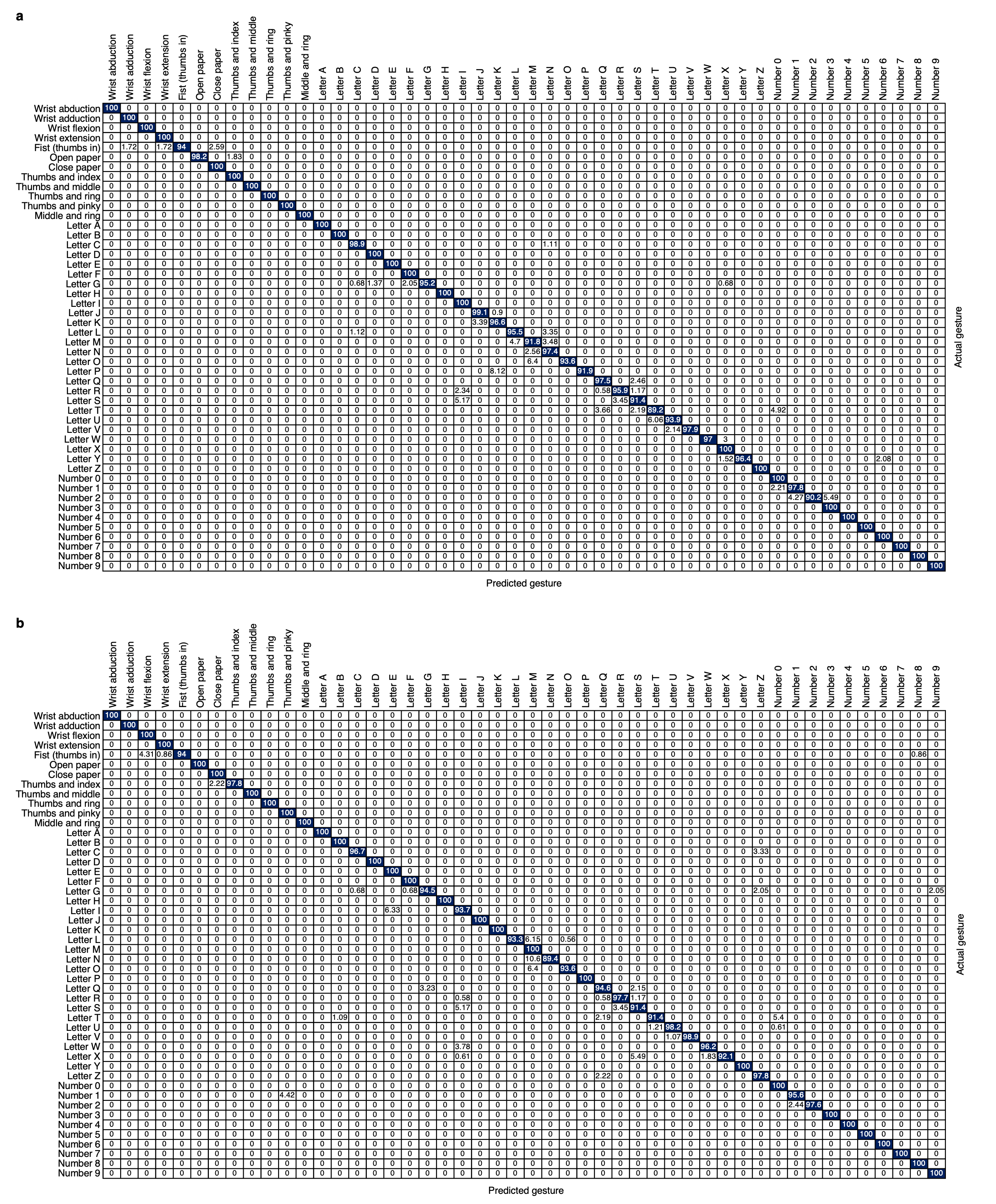}
     \label{fig:supp_5}
     \caption{Inter-session cross-validation confusion matrix for \textbf{a,} subject 3 (accuracy: 97.91 \%), and \textbf{b,} subject 4 (accuracy: 98.01 \%).}
      
\end{figure*}

\section*{Supplementary Data Figure 6. Detailed static gesture recognition - Part 3}

\begin{figure*}[h]
     \centering
     \includegraphics[width=0.7\textwidth]{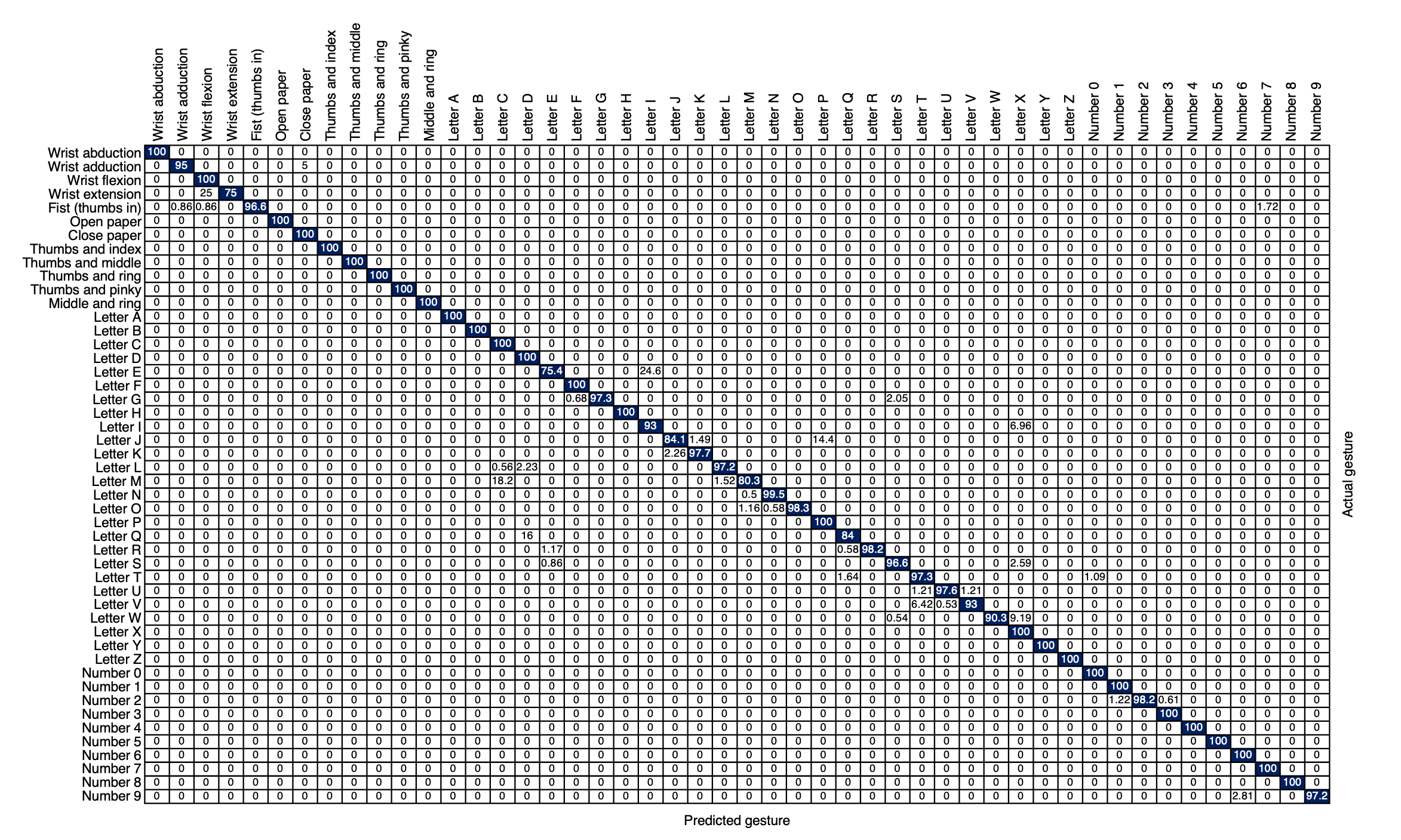}
     \label{fig:supp_6}
     \caption{Inter-session cross-validation confusion matrix for subject 5 (accuracy: 96.70 \%).}

\end{figure*}

\newpage

\section*{Supplementary Data Figure 7. Detailed object recognition - Part 1}

\begin{figure*}[h]
     \centering
     \includegraphics[width=0.9\textwidth]{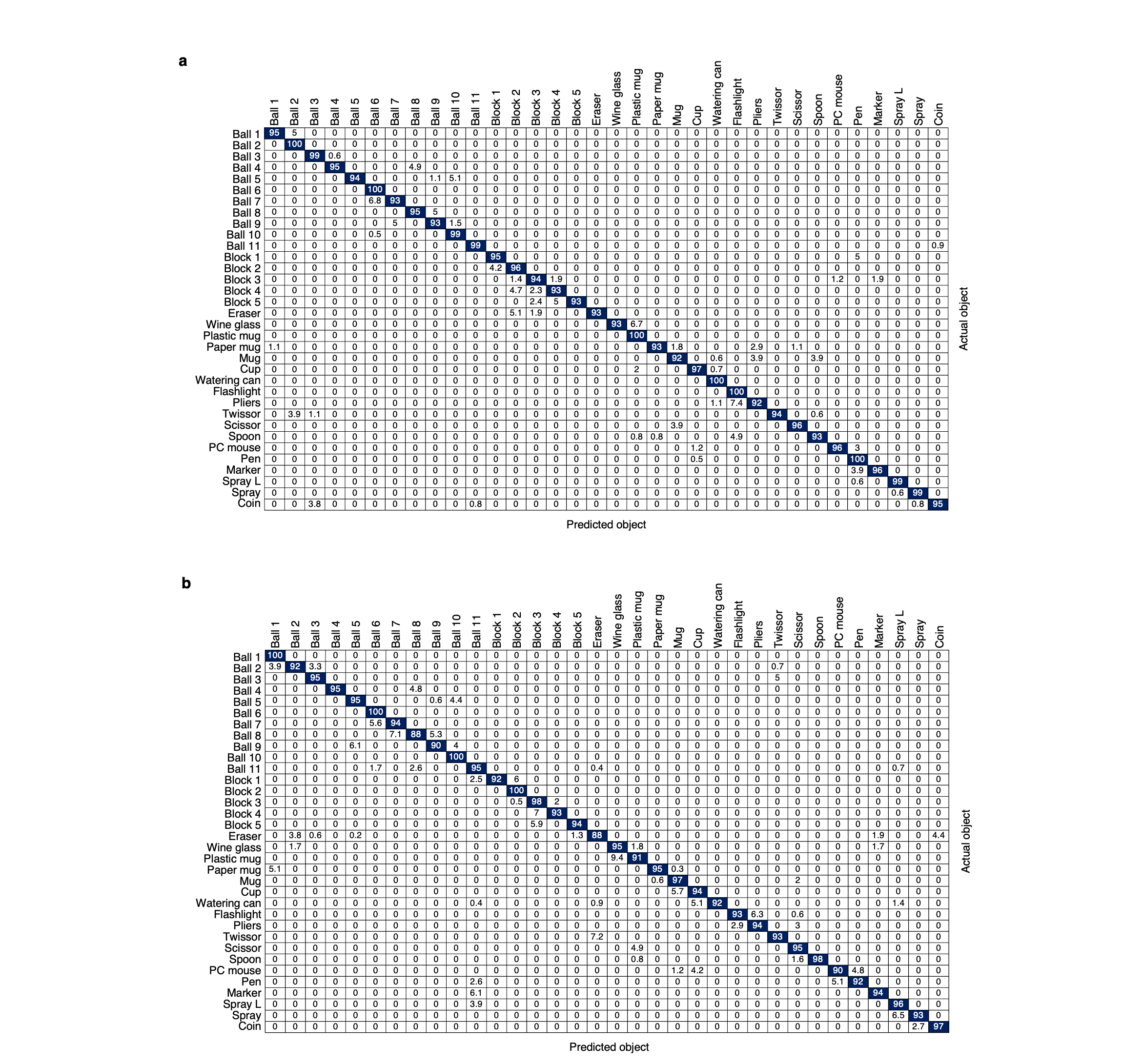}
     \label{fig:supp_7}
     \caption{Inter-session cross-validation confusion matrix for \textbf{a,} subject 1 (accuracy: 95.96 \%), and \textbf{b,} subject 2 (accuracy: 94.38 \%).}
      
\end{figure*}

\newpage

\section*{Supplementary Data Figure 8. Detailed object recognition - Part 2}

\begin{figure*}[h]
     \centering
     \includegraphics[width=0.9\textwidth]{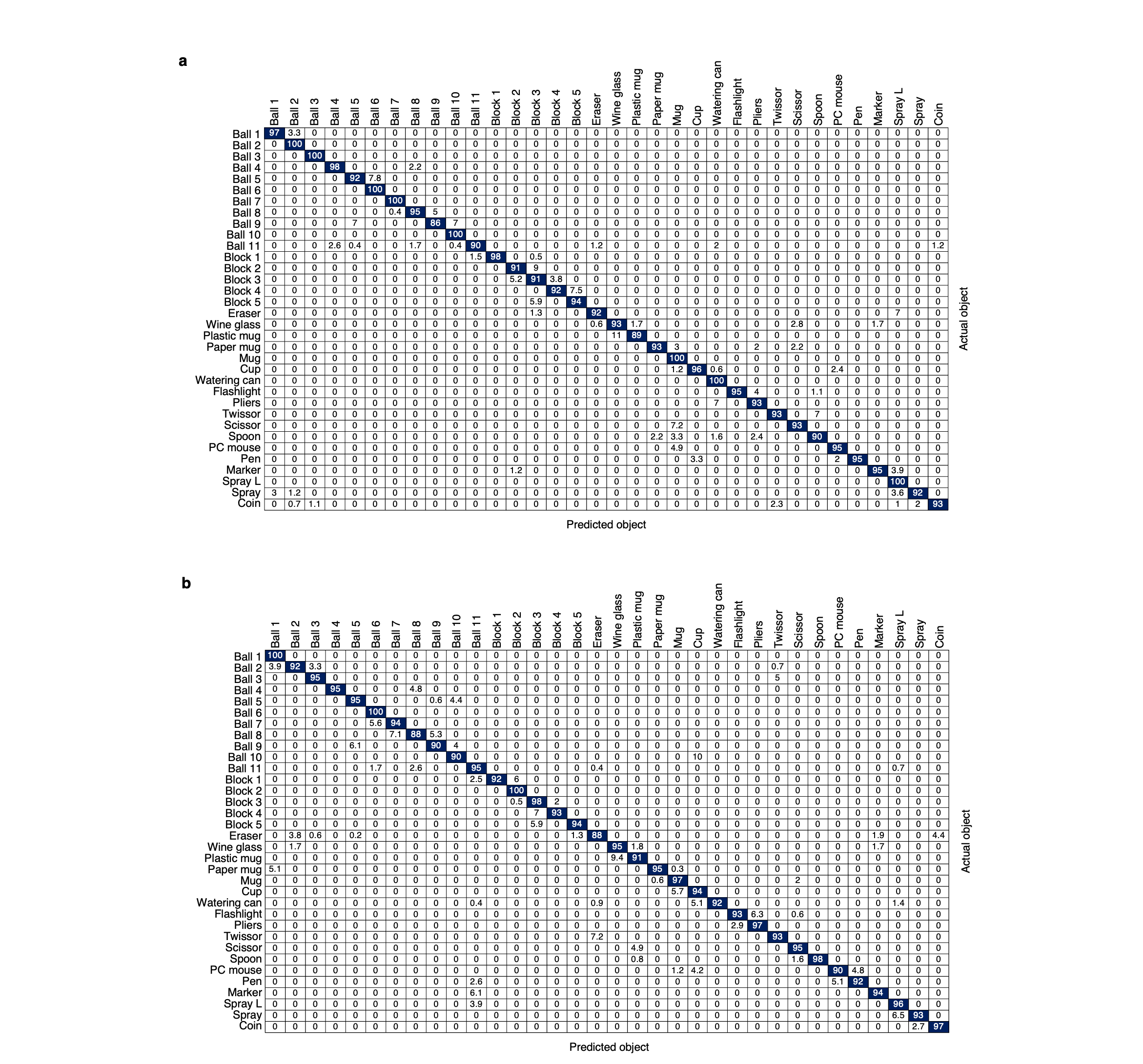}
     \label{fig:supp_8}
     \caption{Inter-session cross-validation confusion matrix for \textbf{a,} subject 3 (accuracy: 94.74 \%), and \textbf{b,} subject 4 (accuracy: 94.17 \%).}
      
\end{figure*}

\newpage

\section*{Supplementary Data Figure 9. Detailed object recognition - Part 3}

\begin{figure*}[h]
     \centering
     \includegraphics[width=0.9\textwidth]{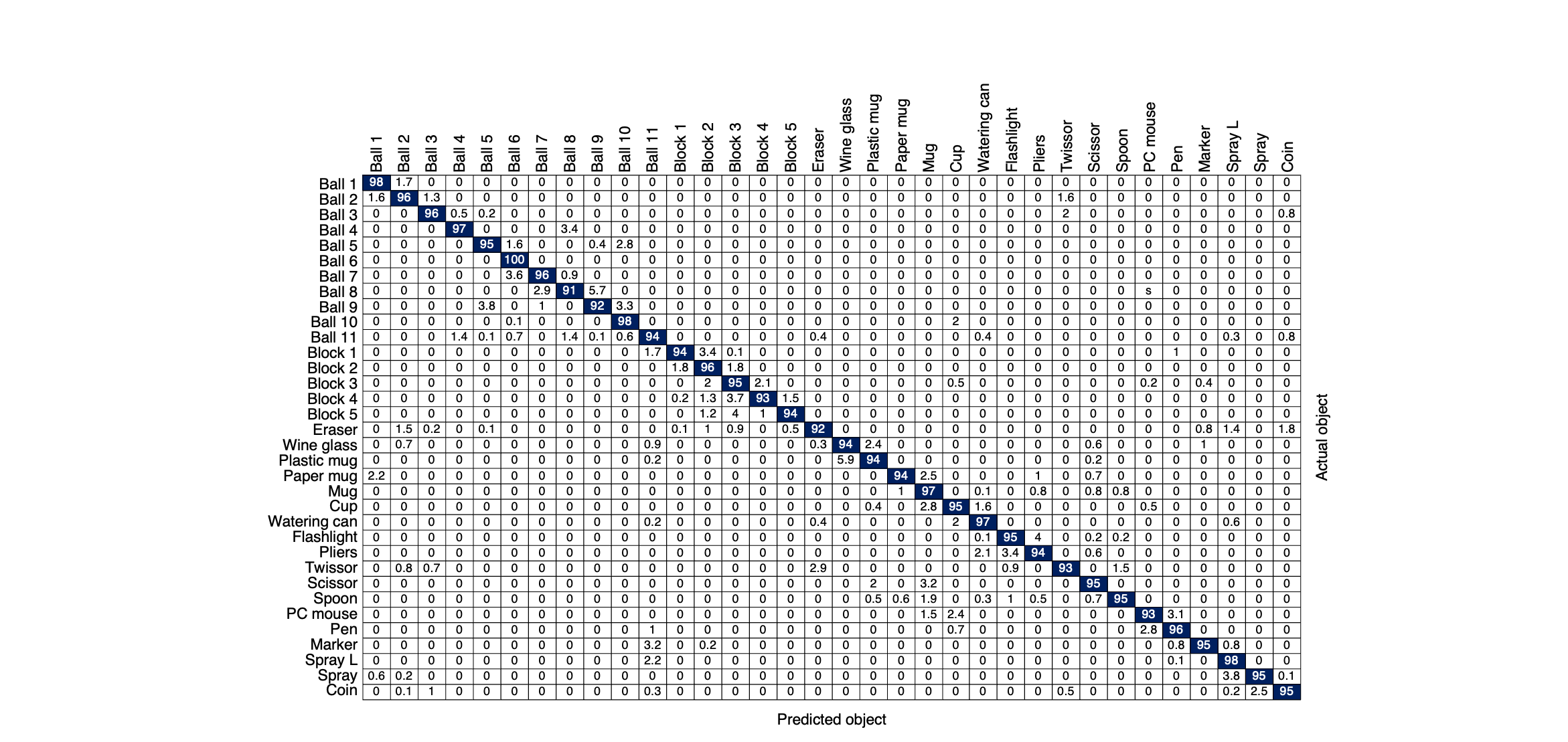}
     \label{fig:supp_9}
     \caption{Inter-session cross-validation confusion matrix for subject 5 (accuracy: 95.83 \%).}
      
\end{figure*}

\newpage

\section*{Supplementary Data Figure 10. Comparison of GlovePoseML and customized GlovePoseML with separate network branches for different modalities.}
\begin{figure*}[h]
     \centering
     \includegraphics[width=0.7\textwidth]{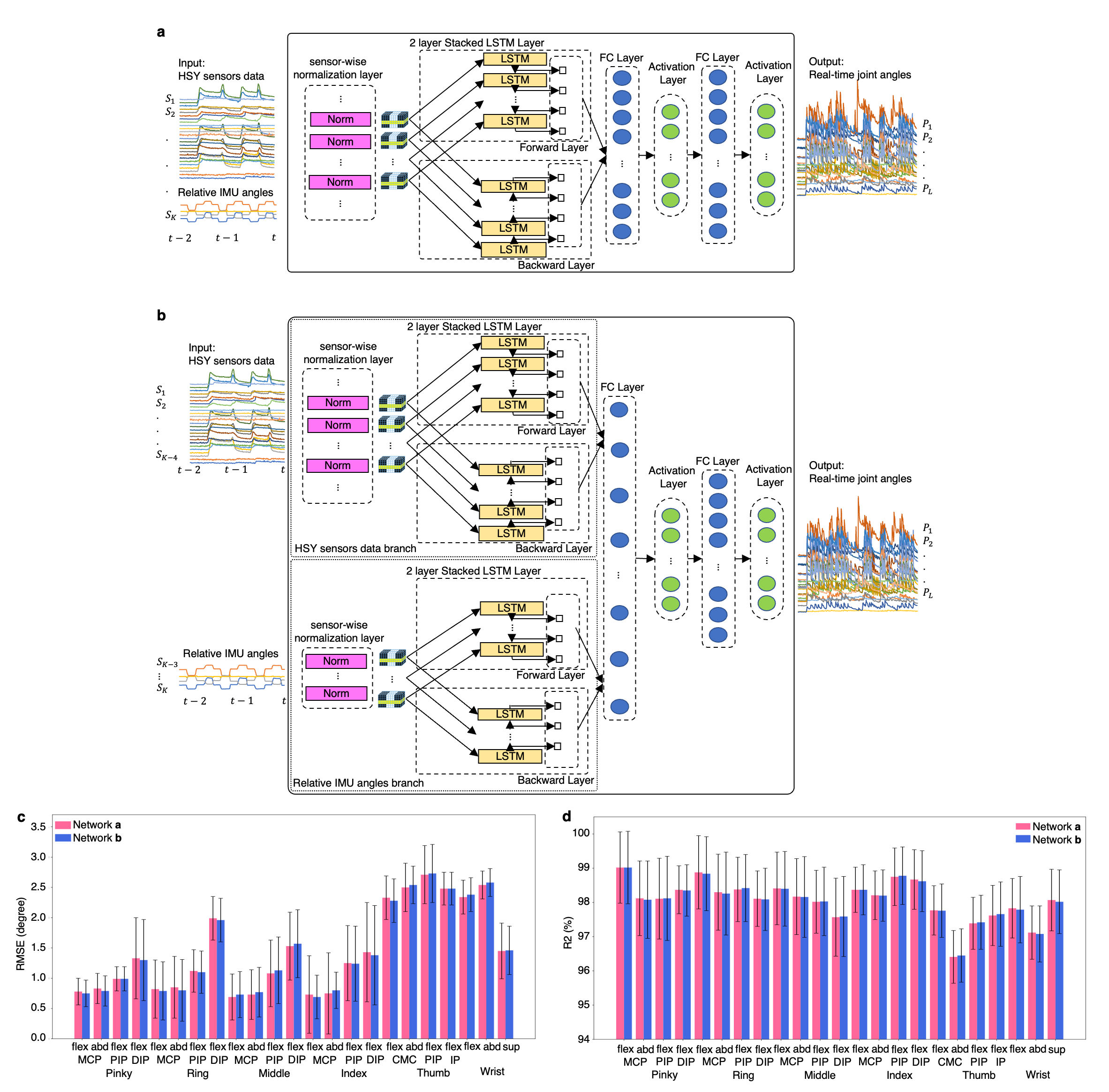}
     \label{fig:supp_10}
     \caption{\textbf{a,} Proposed GlovePoseML network architecture. \textbf{b,} Customized GlovePoseML with different branches for HSY sensors data and relative IMU angle data. Comparision of Network a and Network b in terms of joint-wise \textbf{c,} root mean square error (degree), and \textbf{d,} R2 (\%).}
      
\end{figure*}

\newpage

\section*{Supplementary Data Figure 11. Performance comparison of using different models.}
\begin{figure*}[h]
     \centering
     \includegraphics[width=0.9\textwidth]{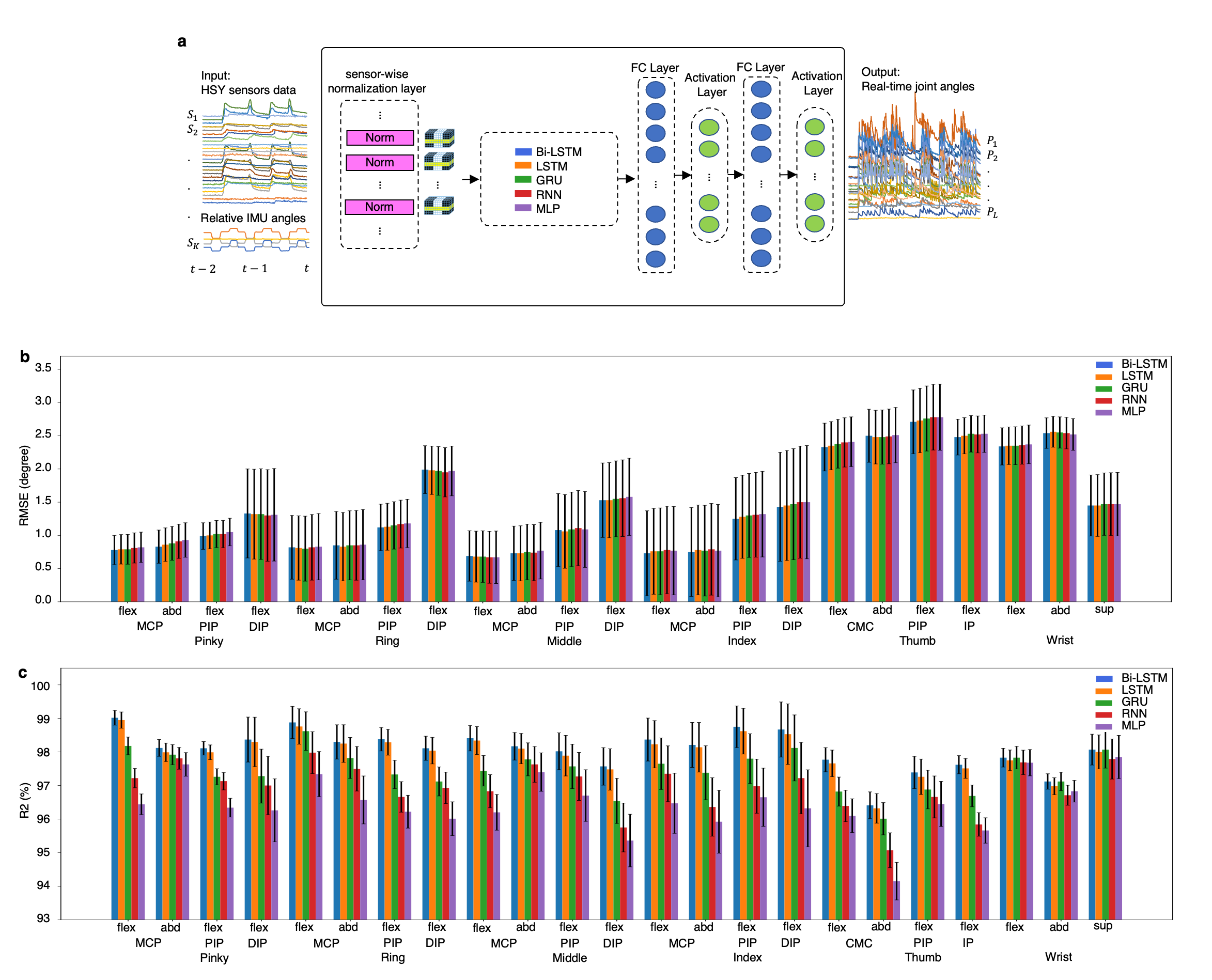}
     \label{fig:supp_11}
     \caption{\textbf{a,} Network architecture used in this study. Joint-wise comparison of networks in terms of \textbf{b,} average root mean square error (degree), and \textbf{c,} R2 (\%). For a fair comparison, we fixed all the corresponding parameters, such as the hidden layer size and the number of epochs between models.}
      
\end{figure*}

\newpage

\section*{Supplementary Data Figure 12. Performance comparison of GlovePoseML trained on normal versus augmented data.}
\begin{figure*}[h]
     \centering
     \includegraphics[width=0.7\textwidth]{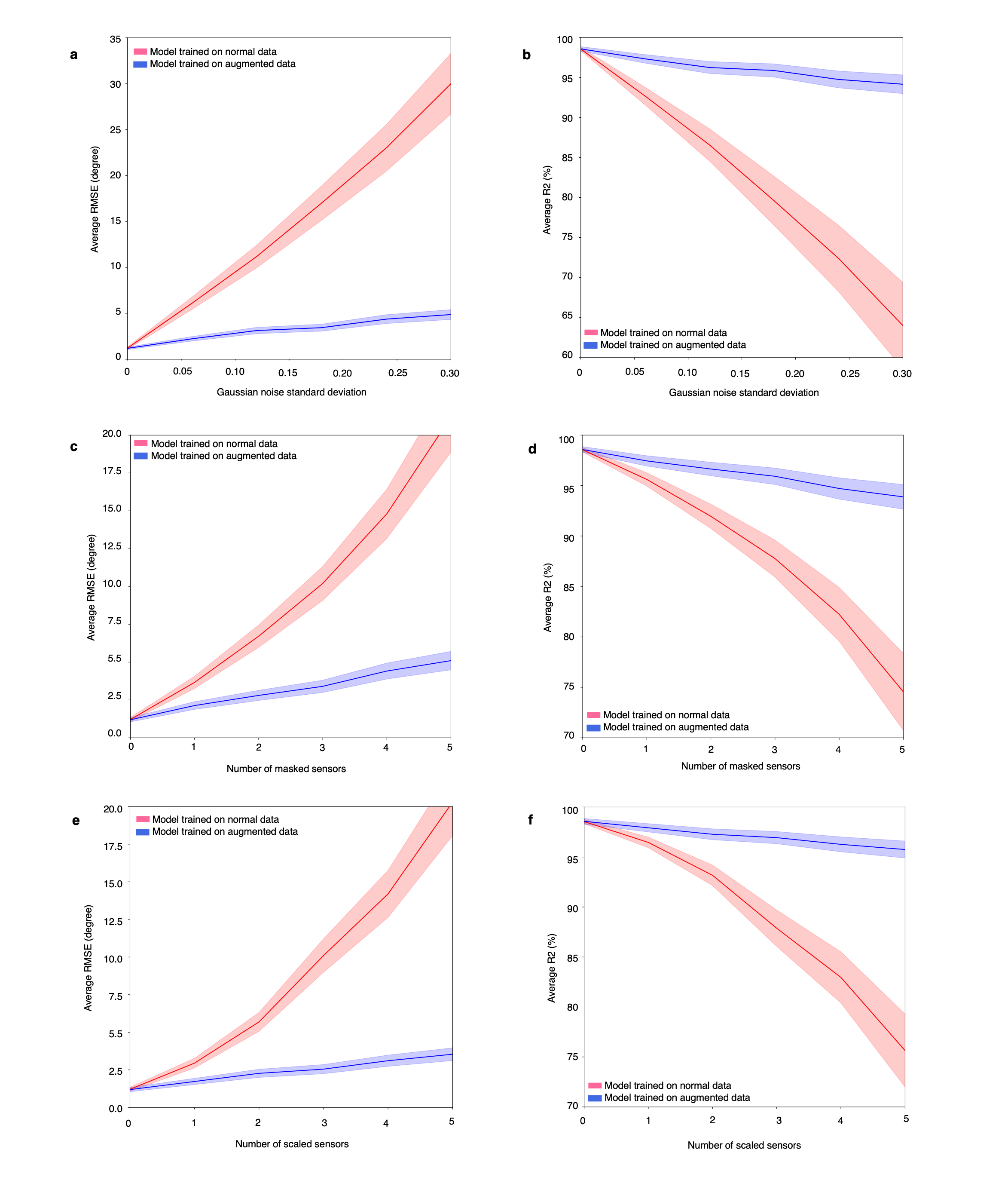}
     \label{fig:supp_12}
     \caption{Effect of adding Gaussian noise to all sensors on model performance in terms of \textbf{a,} average root mean square error (degree), and \textbf{b,} R2 (\%). Effect of randomly masking sensors in terms of \textbf{c,} average   mean square error (degree), and \textbf{d,} R2 (\%). Effect of randomly scaling sensors from 0.5 to 1.5 in terms of \textbf{e,} average root mean square error (degree), and \textbf{f,} R2 (\%).}
      
\end{figure*}

\begin{landscape}
\section*{Supplementary Table 2. Object properties}

\begin{table}[h]
\centering
\resizebox{0.65\columnwidth}{!}{%
\begin{tabular}{|c c c c c c c c c|}
\hline
\multirow{2}{*}{Object type} & \multirow{2}{*}{Object name} & \multirow{2}{*}{Material} & \multirow{2}{*}{Weight {[}g{]}} & \multicolumn{4}{c}{Size {[}cm{]}}                                                               & \multirow{2}{*}{Young's modulus {[}GPa{]}} \\ 
                        &                              &                           &                                 & 
                        \multicolumn{1}{c}{length} & \multicolumn{1}{c}{width} & \multicolumn{1}{c}{depth} & diameter &                                            \\\hline 
\multirow{11}{*}{Balls}                       & Ball 1                       & Rubber                    & 180.5                           & \multicolumn{1}{c}{}       & \multicolumn{1}{c}{}      & \multicolumn{1}{c}{}      & 13.4     & 0.004                                      \\ 
                       & Ball 2                       & Polyester/Foam            & 19.5                            & \multicolumn{1}{c}{}       & \multicolumn{1}{c}{}      & \multicolumn{1}{c}{}      & 11.7     & 0.002                                      \\ 
                       & Ball 3                       & PU/Rubber/Cork            & 139.5                           & \multicolumn{1}{c}{}       & \multicolumn{1}{c}{}      & \multicolumn{1}{c}{}      & 7.3      & 0.550                                      \\ 
                      & Ball 4                       & PU foam                   & 28.5                            & \multicolumn{1}{c}{}       & \multicolumn{1}{c}{}      & \multicolumn{1}{c}{}      & 6.9      & 0.043                                      \\ 
                       & Ball 5                       & Rubber/wool               & 55.0                            & \multicolumn{1}{c}{}       & \multicolumn{1}{c}{}      & \multicolumn{1}{c}{}      & 6.3      & 0.035                                      \\ 
                       & Ball 6                       & Rubber                    & 144.0                           & \multicolumn{1}{c}{}       & \multicolumn{1}{c}{}      & \multicolumn{1}{c}{}      & 6.2      & 0.097                                      \\ 
                       & Ball 7                       & PU foam                   & 12.5                            & \multicolumn{1}{c}{}       & \multicolumn{1}{c}{}      & \multicolumn{1}{c}{}      & 5.8      & 0.015                                      \\ 
                      & Ball 8                       & Plastic                   & 61.5                            & \multicolumn{1}{c}{}       & \multicolumn{1}{c}{}      & \multicolumn{1}{c}{}      & 4.6      & \multirow{4}{*}{Rigid}                                      \\ 
                       & Ball 9                       & Plastic                   & 4.0                             & \multicolumn{1}{c}{}       & \multicolumn{1}{c}{}      & \multicolumn{1}{c}{}      & 3.9      &                                       \\ 
                      & Ball 10                      & Marble                    & 19.0                            & \multicolumn{1}{c}{}       & \multicolumn{1}{c}{}      & \multicolumn{1}{c}{}      & 2.5      &                                       \\ 
                      & Ball 11                      & Marble                    & 5.5                             & \multicolumn{1}{c}{}       & \multicolumn{1}{c}{}      & \multicolumn{1}{c}{}      & 1.6      &                                       \\ \hline
\multirow{6}{*}{Blocks}                      & Block 1                      & Wood                      & 351.0                           & \multicolumn{1}{c}{10.0}   & \multicolumn{1}{c}{10.0}  & \multicolumn{1}{c}{10.0}  &          & \multirow{6}{*}{Rigid}                                      \\ 
                      & Block 2                      & Wood                      & 236.0                           & \multicolumn{1}{c}{7.5}    & \multicolumn{1}{c}{7.5}   & \multicolumn{1}{c}{7.5}   &          &                                       \\ 
                      & Block 3                      & Wood                      & 48.0                            & \multicolumn{1}{c}{5.0}    & \multicolumn{1}{c}{5.0}   & \multicolumn{1}{c}{5.0}   &          &                                       \\
                      & Block 4                      & Wood                      & 6.5                             & \multicolumn{1}{c}{2.5}    & \multicolumn{1}{c}{2.5}   & \multicolumn{1}{c}{2.5}   &          &                                       \\ 
                      & Block 5                      & Ceramic                   & 37.5                            & \multicolumn{1}{c}{10.2}   & \multicolumn{1}{c}{2.5}   & \multicolumn{1}{c}{0.7}   &          &                                       \\ 
                      & Eraser                       & Polystyrene               & 11.0                            & \multicolumn{1}{c}{12.6}   & \multicolumn{1}{c}{5.1}   & \multicolumn{1}{c}{2.6}   &          &                                       \\ \hline
\multirow{6}{*}{Drinkware}                      & Wine glass                   & Plastic                   & 103.0                           & \multicolumn{1}{c}{}       & \multicolumn{1}{c}{}      & \multicolumn{1}{c}{19.6}  & 9.2      & \multirow{6}{*}{Rigid}                                      \\ 
                      & Plastic mug                  & Plastic                   & 70.5                            & \multicolumn{1}{c}{}       & \multicolumn{1}{c}{}      & \multicolumn{1}{c}{12.5}  & 8.5      &                                       \\ 
                      & Paper mug                    & Paper                     & 11.0                            & \multicolumn{1}{c}{}       & \multicolumn{1}{c}{}      & \multicolumn{1}{c}{10.7}  & 7.5      &                                       \\ 
                      & Mug                          & Ceramic                   & 319.0                           & \multicolumn{1}{c}{}       & \multicolumn{1}{c}{}      & \multicolumn{1}{c}{9.7}   & 8.3      &                                       \\ 
                      & Cup                          & PP plastic                & 45.5                            & \multicolumn{1}{c}{}       & \multicolumn{1}{c}{}      & \multicolumn{1}{c}{7.3}   & 9.3      &                                       \\ 
                      & Watering can                 & Plastic                   & 438.0                           & \multicolumn{1}{c}{30.6}   & \multicolumn{1}{c}{11.0}  & \multicolumn{1}{c}{11.0}  &          &                                       \\ \hline
\multirow{11}{*}{Other}                      & Flashlight                   & Metal                     & 139.0                           & \multicolumn{1}{c}{}       & \multicolumn{1}{c}{}      & \multicolumn{1}{c}{13.2}  & 2.9      & \multirow{11}{*}{Rigid}                                      \\ 
                      & Plier                        & Metal                     & 56.0                            & \multicolumn{1}{c}{11.7}   & \multicolumn{1}{c}{6.0}   & \multicolumn{1}{c}{6.6}   &          &                                       \\ 
                      & Tweezers                     & Metal                     & 16.5                            & \multicolumn{1}{c}{12.1}   & \multicolumn{1}{c}{1.0}   & \multicolumn{1}{c}{1.1}   &          &                                       \\ 
                      & Scissors                     & Metal/plastic             & 31.5                            & \multicolumn{1}{c}{15.0}   & \multicolumn{1}{c}{6.0}   & \multicolumn{1}{c}{0.8}   &          &                                       \\ 
                      & Spoon                        & Metal                     & 64.0                            & \multicolumn{1}{c}{20.1}   & \multicolumn{1}{c}{4.5}   & \multicolumn{1}{c}{0.3}   &          &                                       \\ 
                      & PC mouse                     & Plastic/electroncis       & 102.0                           & \multicolumn{1}{c}{10.7}   & \multicolumn{1}{c}{7.4}   & \multicolumn{1}{c}{3.6}   &          &                                       \\ 
                      & Pen                          & Plastic                   & 11.5                            & \multicolumn{1}{c}{}       & \multicolumn{1}{c}{}      & \multicolumn{1}{c}{14.0}  & 1.2      &                                       \\ 
                      & Marker                       & Plastic                   & 14.0                            & \multicolumn{1}{c}{}       & \multicolumn{1}{c}{}      & \multicolumn{1}{c}{12.1}  & 1.7      &                                       \\ 
                      & Spray L                      & Plastic/water             & 142.5                           & \multicolumn{1}{c}{}       & \multicolumn{1}{c}{}      & \multicolumn{1}{c}{16.4}  & 6.5      &                                       \\
                      & Spray                        & Plastic/water             & 125.0                           & \multicolumn{1}{c}{}       & \multicolumn{1}{c}{}      & \multicolumn{1}{c}{13.9}  & 3.8      &                                       \\ 
                      & Coin                         & Metal                     & 2.0                             & \multicolumn{1}{c}{}       & \multicolumn{1}{c}{}      & \multicolumn{1}{c}{0.1}   & 1.8      &                                       \\ \hline
\end{tabular}%
}
\end{table}
\end{landscape}

\section*{Supplementary Video 1}
Dynamic articulated tracking of finger movements.

\vspace{-5pt}

\section*{Supplementary Video 1}
Dynamic articulated tracking of finger movements.

\vspace{-5pt}

\section*{Supplementary Video 2}
Typing on a mock keyboard.

\vspace{-5pt}

\section*{Supplementary Video 3}
3D drawing in air.

\vspace{-5pt}

\section*{Supplementary Video 4}
Static hand gesture recognition.

\vspace{-5pt}

\section*{Supplementary Video 5}
Dynamic hand gesture recognition.

\vspace{-5pt}

\section*{Supplementary Video 6}
Object detection based on the participants' grasp pattern.

\end{document}